\documentclass[twocolumns]{aa} 

\usepackage{graphicx}
\usepackage{txfonts}
\usepackage{hyperref}
\usepackage{multirow}
\usepackage{comment}

\usepackage{natbib}
\begin{document} 

   \title{The structural properties of multiple populations in the dynamically young globular cluster NGC~2419}


   \author{Silvia Onorato \inst{1}
          \and
          Mario Cadelano \inst{2,3}
          \and
          Emanuele Dalessandro \inst{3}
          \and
         Enrico Vesperini \inst{4}
          \and
          Barbara Lanzoni \inst{2,3}
          \and
         Alessio Mucciarelli \inst{2,3}
          }

   \institute{Leiden Observatory, Leiden University, PO Box 9513, NL-2300 RA Leiden, The Netherlands\\
         \and
              Dipartimento di Fisica e Astronomia,
              Via Gobetti 93/2 I-40129 Bologna, Italy\\
         \and
             INAF-Astrophysics and Space Science Observatory Bologna,
             Via Gobetti 93/3 I-40129 Bologna, Italy\\
        \and
            Department of Astronomy, Indiana University,
            Swain West, 727 E. 3rd Street, IN 47405 Bloomington, USA\\
             }


 
  \abstract
{NGC~2419 is likely the globular cluster (GC) with the lowest dynamical age in the Galaxy. 
This makes it an extremely interesting target for studying the properties of its multiple populations (MPs), as they have been likely affected only modestly by long-term dynamical evolution effects.
Here we present for the first time a detailed analysis of the structural and morphological properties of the MPs along the whole extension of this remote and massive GC by combining high-resolution HST and wide-field ground-based data.
In agreement with formation models predicting that second population (SP) stars form in the inner regions of the first population (FP) system, we find that the SP is more centrally concentrated than the FP. This may provide constraints on the relative concentrations of MPs in the cluster early stages of the evolutionary phase driven by two-body relaxation. In addition, we find that the fraction of FP stars is larger than expected from the
general trend drawn by Galactic GCs. If, as suggested by a number of studies, NGC~2419 formed in the Sagittarius dwarf galaxy and was later accreted by the Milky Way, we show that the observed FP fraction may be explained as due to the transition of NGC~2419 to a weaker tidal field (its current Galactocentric distance is $d_{gc}\sim 95$ kpc) and consequently to a reduced loss rate of FP stars.}

\titlerunning{Multiple Populations in NGC~2419}

\maketitle
%
\section{Introduction}
The presence of sub-populations differing in terms of their light-element abundances (e.g. He, C, N, O, Na, Mg, Al - elements involved in the CNO-cycle reactions) while having the same iron (and iron-peak) content (hereafter multiple stellar populations - MPs) is a key general property of globular clusters (GCs; see e.g. \citealt{Bastian2018} and \citealt{gratton19} for recent reviews). 

\noindent
Stars sharing the light-element abundances of the surrounding field stars (i.e. Na-poor/O-rich, CN-weak) are commonly classified as first-population (FP), while Na-rich/O-poor, CN-strong stars are referred to as second-population (SP).
Interestingly, MPs can be also studied through photometry exploiting specific combinations of UV and optical filters (e.g. \citealt{Sbordone2011}). In fact, variations of light-element abundances among stars can result in splitting or spreads of different evolutionary sequences in the color-magnitude diagram (CMD) when appropriate filter combinations
are adopted (e.g. \citealt{piotto15,larsen2015,Milone2017,Lee2019}). The efficiency of these filters combinations in separating sub-populations mostly depends on their sensitivity to variations of the CN and NH molecular abundances.

MPs are observed in nearly all relatively massive ($M>10^4 M_{\odot}$; \citealt{Carretta2010}) stellar clusters, both in the Milky Way and in external galaxies (e.g. \citealt{mucciarelli08,Dalessandro2016,larsen14,sills19}) and in clusters of ages at least down to $\sim$ 1.5 Gyr \citep{Martocchia2018,cadelano2022}\footnote{It is worth noticing, however, that the detection in younger systems seems to be limited to the main sequence only \citep{cadelano2022}. }.

MPs are believed to form during the very early epochs of GC life ($\sim 10-100$ Myr). A number of theoretical studies have been put forward over the years, but no consensus has been reached yet on their origin (e.g.  \citealt{Decressin2007a,dercole08,Bastian2013,Denissenkov2014,dantona16,gieles18}). It is now clear that a comprehensive approach able to focus on different aspects of GC formation, including chemical and dynamical evolution, is the key to drive a significant progress in this field.

\noindent
The kinematical and structural properties of MPs can provide important insights into the early epochs of GC formation and evolution. One of the predictions of MPs formation models (see e.g. \citealt{dercole08,calura19}) is that SP stars form a centrally segregated sub-system possibly 
characterized by a more rapid internal rotation \citep{Bekki2011,lacchin22} than the more spatially extended FP system. Although the original structural and kinematical differences between FP and SP stars are gradually erased during GC long-term dynamical evolution (see e.g. \citealt{Vesperini2013,Henault-Brunet2015,miholics15}), some clusters are expected to retain some memory of their primordial properties. 
In fact, sparse and inhomogeneous observations show that MPs are characterized by (quite) remarkable differences in their relative structural parameters/radial distributions (e.g., \citealt{Lardo2011,Dalessandro2016,Dalessandro2018,Simioni2016}), different degrees of orbital anisotropy (e.g., \citealt{Richer2013,Bellini2015,libralato23}), different rotation amplitudes (e.g., \citealt{Lee2015,Cordero2017,dalessandro2021}) and significantly different binary fractions \citep{Lucatello2015,Dalessandro2018a,kamann2020}.

However, despite the key importance of a full understanding of the structural/kinematical properties of MPs, no systematic study has been performed so far and a complete picture is still lacking. 
In \citet{Dalessandro2019} (see also \citealt{leitinger23}), we have carried out a first step in this direction and studied the spatial distributions of MPs in a sample of 20 GCs spanning a broad range of dynamical ages. The differences between the spatial distributions of FP and SP stars were quantitatively measured by means of the parameter $A^+$, defined as the area enclosed between their cumulative radial distributions. Our study has revealed a clear trend between $A^+$ and GC dynamical evolution, as constrained by the ratio between the cluster age and its half-mass relaxation time (t/t$_{rh}$). Less dynamically evolved clusters (t/t$_{rh} < 8 - 10$) have SP stars more centrally concentrated than FPs (i.e. negative values of $A^+$), while in more dynamically evolved systems the spatial differences between FP and SP stars decrease and eventually disappear ($A^+$ tends to zero). This is the first purely observational evidence of the dynamical path followed by the structural properties of MPs toward a complete FP-SP spatial mixing. Such a behavior is consistent with predictions by $N$-body and Monte Carlo simulations following the long-term dynamical evolution of MPs \citep{Vesperini2013,Vesperini2018,Dalessandro2019,Vesperini2021,Sollima2021} in clusters forming with an initially more centrally concentrated SP sub-system. Very interestingly, the existence of such an evolutionary path provides the possibility to trace back the structural properties of MPs before two-body relaxation and other long-term dynamical processes alter the cluster's structural properties. These constraints, ideally in combination with information on the very early structural properties of proto-clusters (possibly obtained with JWST observations; see e.g. \citealt{vanzella2017}) are critical to test existing cluster formation and evolutionary models and guide the development of new formation scenarios.

In this context, the structural properties of dynamically young systems (i.e. those with $t/t_{rh}< 3 - 4$) are particularly meaningful, as they are expected to be only partially affected by long-term dynamical evolution and therefore retain better memory of the conditions emerging from the formation and early evolutionary phases. As a consequence, the study of dynamically young clusters can allow us to probe the early structural properties of MPs, thus better defining their dynamical evolutionary path. The key constraints provided by this analysis have also important implications on the interpretation of the other MP kinematical features that are becoming observable now thanks to Gaia and the high multiplexing capabilities of state of the art multi-object  and integral field unit (IFU) spectrographs.
However, in our initial study \citep{Dalessandro2019}, only three clusters fall in this critical range of young dynamical ages, thus preventing any meaningful exploration of this kind.

In this paper we analyse in detail the case of NGC~2419. This is an old (t$\sim12$ Gyr; \citealt{Dalessandro2008}) and metal-poor ([Fe/H]=-2.1; \citealt{Mucciarelli2012}) Galactic GC, and it shows quite extreme light-element chemical patterns. In fact, about half of the stars in NGC~2419 has extremely depleted Mg abundances, down to [Mg/Fe]$\sim-1$,\footnote{It is worth noticing that Mg depleted stars in GCs typically reach values down to [Mg/Fe]$\sim-0.4$ \citep[e.g.][]{meszaros2021,alvarez22}, thus making NGC~2419 a peculiar and extreme case.} along with strong enrichment in K \citep{cohen12,Mucciarelli2012}.  In addition, NGC~2419 hosts a sub-population ($\sim30\%$) of extremely He-rich stars ($Y\sim0.40$), as constrained through photometry of red giant branch (RGB) stars  and through the cluster horizontal branch (HB) morphology \citep{dicriscienzo11}. 
More interestingly in this context, NGC~2419 is one of the most massive (M$\sim10^6 M_{\odot}$; \citealt{Baumgardt2018})  
GC in the Galaxy and its relaxation time is among the largest ($t_{rh}\sim 42.7$ Gyr; \citealt{Harris2010}).
In addition, with a Galactocentric distance d$_{GC}\sim95$ kpc, NGC~2419 is expected to be 
only marginally affected by gravitational interactions with the Galactic potential that can alter its dynamical properties.
The very young dynamical age of NGC~2419 has been also empirically confirmed by \citet[][see also \citealt{ferraro2018}]{Dalessandro2008} and \citet{Bellazzini2012} from to the analysis of the cluster blue straggler star population and mass-function slope radial variations, respectively. 
It is also interesting to note that, thanks to the kinematic information secured by Gaia, it has been suggested that NGC~2419 was likely born in the Sagittarius dwarf galaxy and that it was later accreted by the Milky Way (e.g., \citealt{massari19,bellazzini2020}) . 

First analyses of the radial distribution of MPs in NGC~2419 were presented by \citet{Beccari2013}, who found that SP stars are significantly more centrally concentrated than the FP sub-population, based on wide-field ground-based Large Binocular Telescope (LBT) data limited mostly to the external cluster regions ($>100\arcsec$), and by \citet{larsen19} who used Hubble Space Telescope (HST) observations limited only to the innermost $80\arcsec$ and found (with low significance) that SP stars are only slightly more centrally concentrated than the FP.

To perform a complete and detailed analysis of the MP radial distribution in NGC~2419, in this paper we combine high-resolution HST photometry to sample the innermost and more crowded cluster regions and wide-field images to adequately sample the entire cluster extension. 

The paper is structured as follows. The observations and data analysis procedures are detailed in Section~2. In Section 3 we describe the selection of the different sub-populations. In Section 4 we estimate the structural properties of MPs, analyse their cumulative radial distributions, and compare them with the results of \citet{Dalessandro2019}. We summarize our conclusions in Section~5.


\section{Observations and Data Analysis}
\label{sec:obs}
This work is based on a combination of $22$ images obtained with the HST WFC3/UVIS camera through the F336W and F343N filters (GO-15078, PI: Larsen), and the F438W and F814W filters (GO-11903, PI: Kalirai). A detailed observation log is reported in Table~\ref{tab:dataset}.

\begin{table}[h!]
\caption{Summary of the HST data-set used in this work.}
\label{tab:dataset}
\begin{tabular}{ccc}
\hline \hline
Proposal ID / PI & Filter & $t_{exp}$ (s)\\
\hline
\multirow{2}{9em}{GO-15078 / Larsen} & $F336W$ & $2 \times 1392 + 4 \times 1448$\\
 & $F343N$ & $4 \times 1392 + 8 \times 1448$\\ \hline
\multirow{2}{9em}{GO-11903 / Kalirai}  & $F438W$ & $2 \times 725$\\
 & $F814W$ & $2 \times 650$\\
\hline
\end{tabular}
\end{table}

The data-reduction was performed on the \texttt{\textunderscore flc} images, which are flat-fielded, bias-subtracted and corrected for dark current and charge transfer efficiency by the Space Telescope Science Institute WFC3 pipeline. The most updated pixel-area-maps were applied independently to each chip and image. 

The photometric analysis was performed independently on each chip by using \texttt{DAOPHOT IV} \citep{Stetson1987}. Few hundreds of bright and isolated stars were selected in each frame to model the point-spread-function (PSF). By following the approach already adopted in previous works of our group (e.g. \citealt{Dalessandro2018,cadelano20psr,cadelano20_m15m30,cadelano20_n6256}) 
a first star list was obtained for each image by independently fitting stellar-like sources above the $4 \sigma$ level from the local background.
We then created a master list composed of stars detected in at least half of the $F343N$ and $F438W$ images. At the corresponding position of stars in this final master-list, a fit was forced with \texttt{DAOPHOT/ALLFRAME} \citep{Stetson1994} in each frame of the data-set. For each star thus recovered, multiple magnitude estimates obtained in each chip were homogenized by using \texttt{DAOMATCH} and \texttt{DAOMASTER}, finally obtaining the final stellar magnitudes and relative uncertainties.

Instrumental magnitudes were reported to the VEGAMAG photometric system using the equations and zero-points reported in the dedicated HST web pages. 
Instrumental coordinates were corrected for geometric distortions by using the prescriptions by \citet{Bohlin2016}. Then, they were reported onto the absolute system ($\alpha$, $\delta$) using first the stars in common with the ACS HST high-resolution catalogue presented by \citet{Dalessandro2008}, and then the stars in common with Gaia DR3 \citep{gaia_dr3} as primary astrometric standards. Catalogues cross-match and geometric transformations were obtained by using the cross-correlation tool \texttt{CataXcorr}\footnote{CataXcorr is a software designed to cross-correlate catalogues of stars in order to search for astrometric solutions; it operates using common sources among the catalogues included. It was developed by Paolo Montegriffo at the INAF-Osservatorio Astronomico di Bologna.}. 

In order to study the properties of the cluster MPs throughout its entire radial extension, we complemented the HST data-set with the wide-field catalogue presented by \citet{Beccari2013} and obtained from observations acquired with the LBC camera mounted at the LBT {using V, I and a SDSS u passbands} (see \citealt{Beccari2013} for details about the data-set and data-analysis). Also in this case, stellar positions were reported to the Gaia DR3 astrometric system to secure homogeneity with the HST data-set. 

\begin{figure}[ht]
\centering
 \includegraphics[scale=0.5]{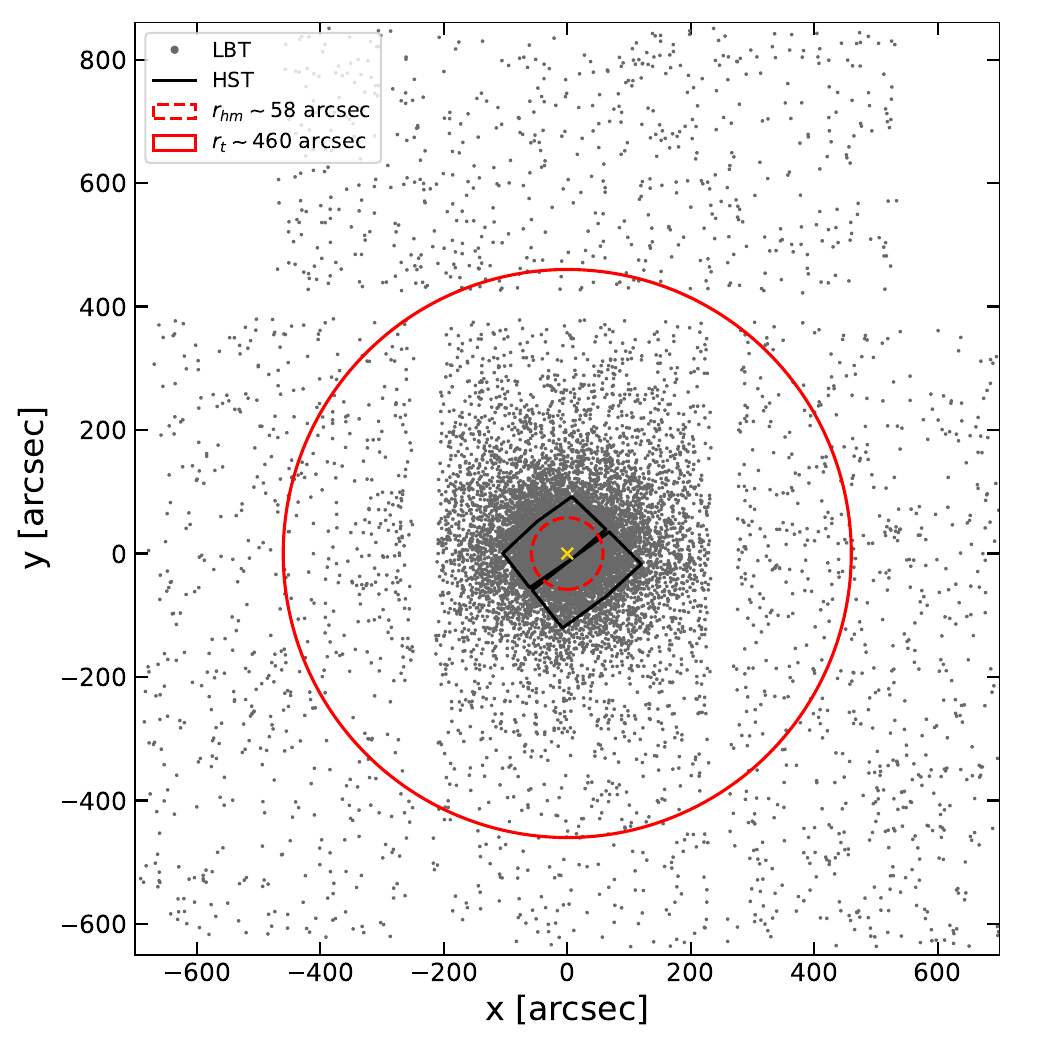}
  \caption{Sky distribution of the surveyed stars measured with respect to the cluster gravitational center. Grey points are stars from the HST adopted data-set and the LBT wide-field catalogue by \citet{Beccari2013}. The black polygon marks the boundaries of the HST FoV. The golden cross marks the cluster gravitational center while the dashed and solid circles correspond to the cluster half-mass and tidal radius, respectively. 
}
 \label{fig:fovtot}
\end{figure}

The sky distribution map of the stars surveyed within the field of view (FoV) of the HST and LBT data-sets is shown in Fig.~\ref{fig:fovtot}. It is worth stressing that the Gaia DR3 catalogue cannot be used to firmly distinguish cluster members from field interlopers along the cluster evolutionary sequences due to the large proper motion uncertainties for such a remote stellar system. However, we statistically estimated that the contamination due to field interlopers is negligible \citep[see e.g.][]{Dalessandro2019_n2173}. In fact, the density of stars in the region beyond the cluster tidal radius and located in the CMD region occupied by the cluster red giant branch is of only $\sim 4\times10^{-5}$ $stars/arcsec^2$, corresponding to only $\sim30$ contaminating stars (out of $\sim2500$) within the cluster tidal radius.

\section{Multiple Populations properties}
\subsection{Multiple Populations selection}

\begin{figure}[ht]
 \centering
 \includegraphics[scale=0.4]{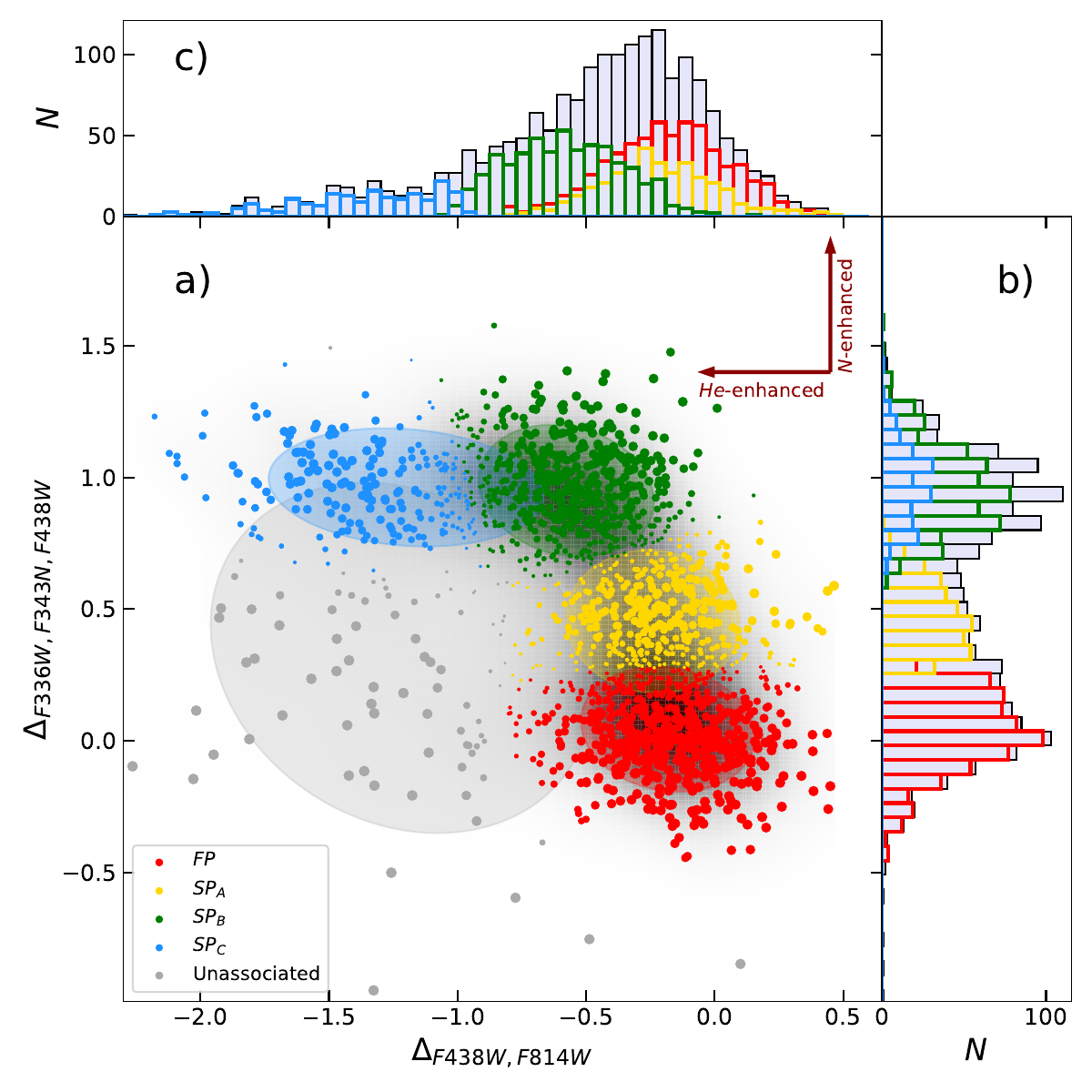}
 \caption{Chromosome map of NGC~2419 RGB stars. \textit{Panel a)}: The different populations as selected through the GMM algorithm are highlighted with different colors. A larger marker size indicates an higher probability that a given star belongs to the assigned population. The confidence ellipsoids of each sub-population fitted with the GMM algorithm are also plotted.  \textit{Panel b)}: Histograms of the distribution along $\Delta_{F336W,F343N,F438W}$ of the total population and of each sub-population. \textit{Panel c)}: Same as in panel b), but for the distribution along $\Delta_{F438W,F814W}$.}
 \label{fig:chm}
\end{figure}

MPs were first selected in the HST data-set along the cluster RGB in the ($m_{F814W}, C_{F336W,F343N,F438W}$) and ($m_{F814W}, m_{F438W} - m_{F814W}$) diagrams, where $C_{F336W,F343N,F438W} = (m_{F336W}-m_{F343N})-(m_{F343N}-m_{F438W})$ (see \citealt{Milone2017}). To this aim, we verticalized the distribution of RGB in the ($m_{F814W}, C_{F336W,F343N,F438W}$) and ($m_{F814W}, m_{F438W} - m_{F814W}$) diagrams with respect to two fiducial lines at the blue and red edges of the RGB in both the CMDs.  The combination of the two verticalized distributions
($\Delta_{F438W,F814W}$ and $\Delta_{F336W,F343N,F438W}$ ) finally gives the so-called cluster Chromosome Map (ChM; Fig.~\ref{fig:chm}). {To maximize the accuracy of the MP selection, only stars with $m_{F814W}<21$ were included in the ChM}. The ChM here obtained is qualitatively similar to the one obtained by \citet{Zennaro2019} and is characterized by a prominent sequence with a quite structured and clumpy stellar distribution and a very sparse and poorly populated sub-group at bluer colors. As shown by the two histograms in Fig.~\ref{fig:chm}, the distribution of $\Delta_{F336W,F343N,F438W}$ is at least bi-modal, 
while the distribution of $\Delta_{F438W,F814W}$ is dominated by a main peak at color $\sim-0.2$ followed by a long tail.

\begin{figure*}[ht]
\centering
 \includegraphics[scale=0.35]{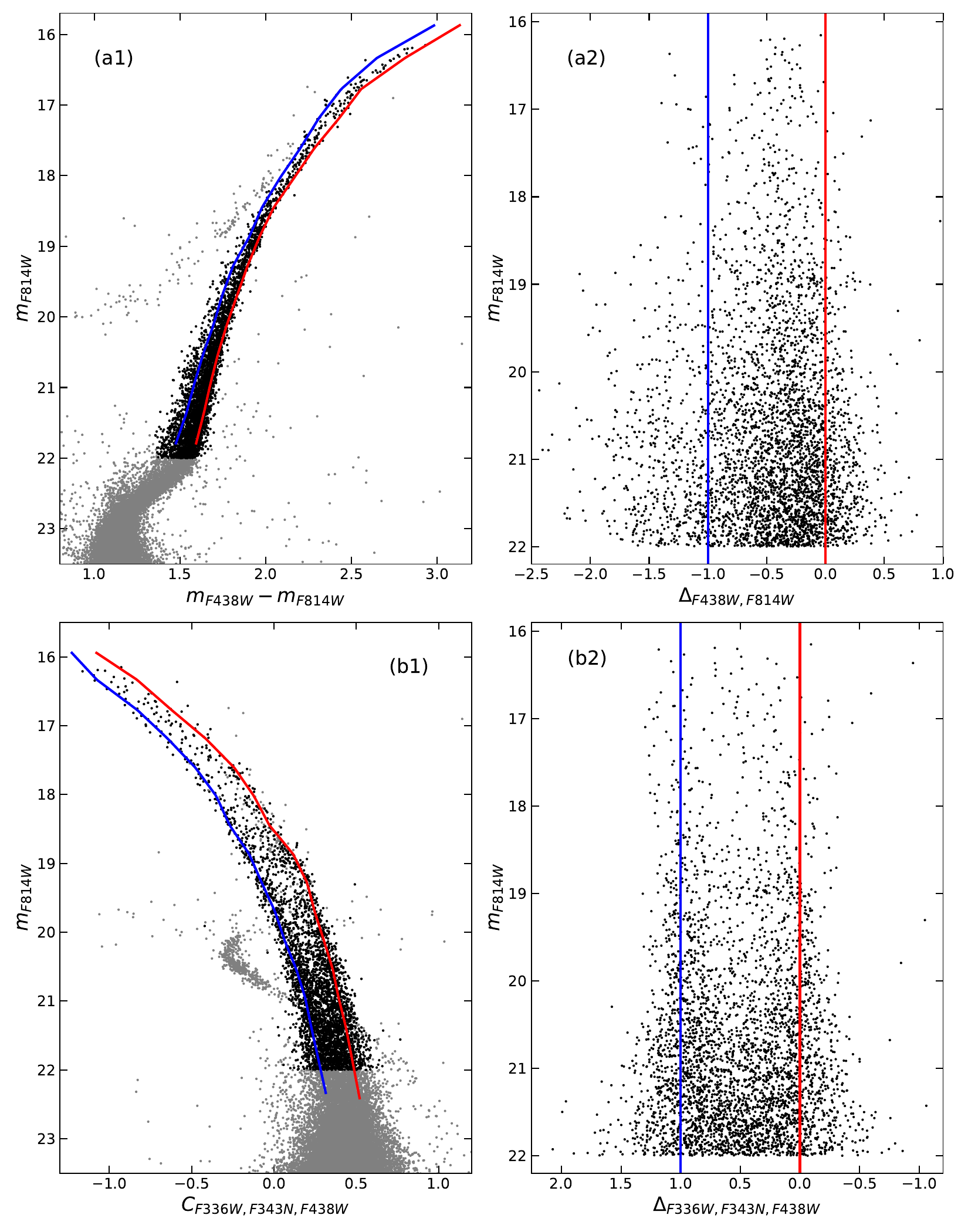}
  \caption{CMDs of NGC~2419 and verticalized RGB color distribution obtained through the HST observations. \textit{Panel a1)}: gray dots are the ($m_{F438W}$, $m_{F438W}-m_{F814W}$) optical CMD, black dots are the selected RGB stars. The red and blue curves are the fiducial lines adopted to verticalize the RGB sequence.  \textit{Panel a2)}: verticalized color distribution of the selected RGB stars. The red and blue vertical lines correspond to the red and blue curves in panel a1). \textit{Panel b1)}: same as in panel a1) but for the ($m_{F814W}, C_{F336W,F343N,F438W}$) CMD. \textit{Panel b2)}: same as in panel a2) but for the filter combination used in panel b1).}
 \label{fig:rlpscol}
\end{figure*}

To identify the number of sub-populations in the cluster and to assign the surveyed stars to the different sub-groups, we fit the ChM with 2D Gaussian mixture models using the \texttt{scikit-learn package}\footnote{\url{https://scikit-learn.org/stable/index.html}} \citep{scikit-learn}. First, we used both the Akaike (AIC) and the Bayes (BIC) Information criteria to infer the optimal number of sub-populations in which the observed ChM can be split. The distributions of both parameters as a function of the number of components reaches a minimum for 5 components, indicating this is the most likely number of sub-groups. We then fit the ChM with a mixture of 5 bi-dimensional Gaussian functions, whose parameters were obtained from the mixture model analysis.  The derived separation among different sub-populations is illustrated by the different colors shown in Figure~\ref{fig:chm}, where the size of each circle corresponds to the probability of each star to belong to a given sub-group. 
In agreement with the findings by \citet{Zennaro2019}, our analysis confirms the presence of a main primordial population centered around the origin of the ChM  (FP - red circles), an intermediate mildly N-enhanced SP sub-group (SP$_A$ - yellow circles), a more strongly N-enriched population (SP$_B$ - green circles) and finally an extreme population significantly enriched both in terms of He and N (SP$_C$ - cyan circles). 
The fifth sparse component is populated by sources of unidentified nature (gray points). This unassociated population might include binaries, evolved blue straggler stars and other exotic objects \citep{Kamann2020_n3201,marino2020} and it will not be considered in the following analysis. 

\begin{figure}
 \centering
 \includegraphics[scale=0.35]{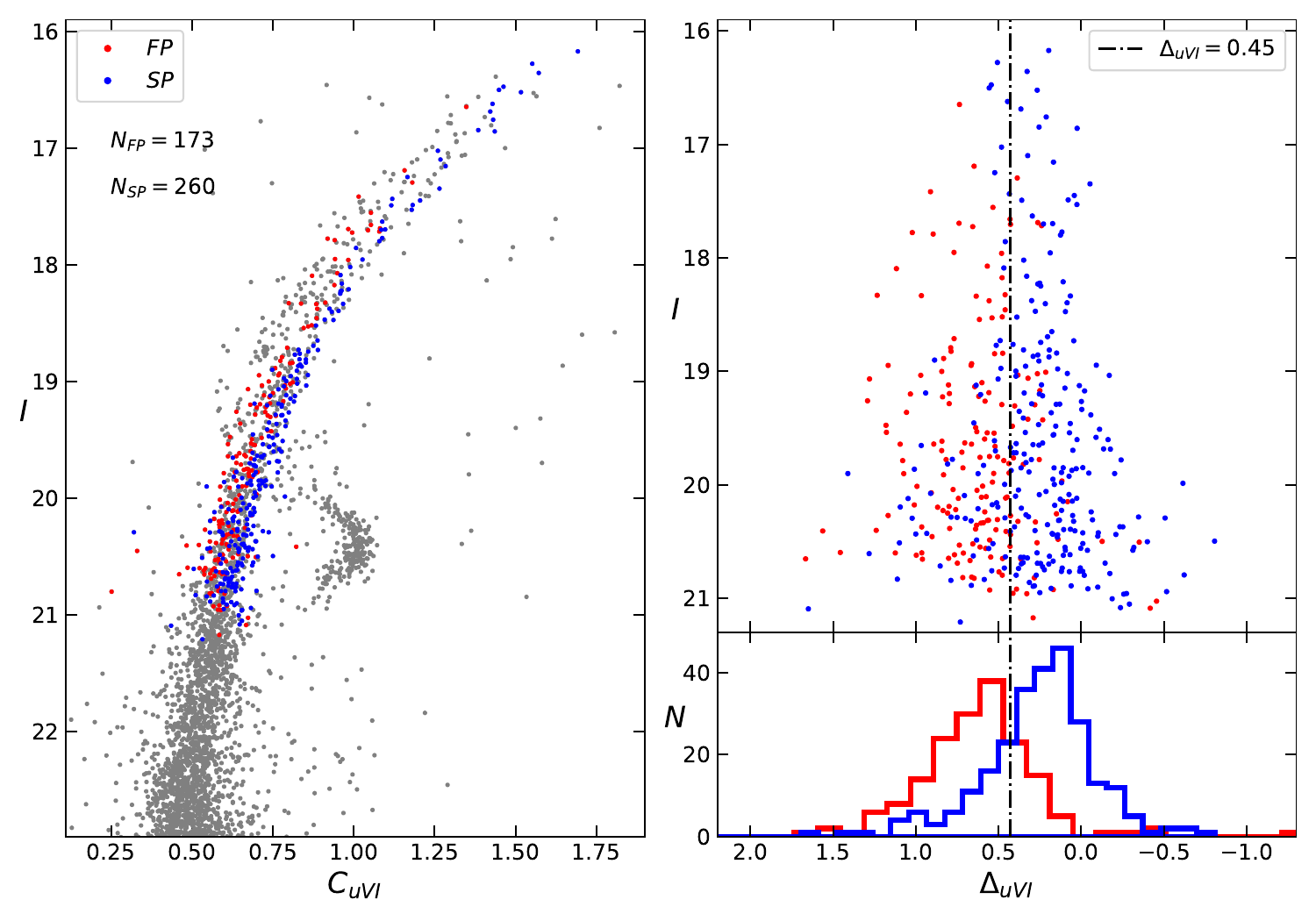}
 \caption{CMDs of NGC~2419 and verticalized RGB color distribution obtained through the LBT observations. {\it Left-hand panel:} $I,C_{uVI}$ CMD of NGC~2419 from the LBT stars in common with the HST catalogue. Red and blue dots are objects classified in the HST data-set as FP and SP stars, respectively. {\it Right-hand panel:} The top panel shows the verticalized distribution of the $C_{uVI}$ color. Red and blue dots are the same as in the left-hand panel. The vertical line is adopted as separation between FP and SP for the LBT stars observed in the area complementary to the HST FoV. {The bottom panel shows the histograms of the verticalized $C_{uVI}$ distributions for FP and SP stars.}}
 \label{fig:cmdBeccari}
\end{figure}

To get a comprehensive view on the MPs structural properties throughout the entire cluster radial extension, we need to identify MPs also in the LBT wide-field catalogue.
 {First of all, we verified the completeness level of the LBT catalogue.  We counted the number of RGB stars in an overlapping area between the HST and LBT dataset and at distances larger than $70\arcsec$ from the cluster center. At the faint end of the RGB analysed in this work ($20<m_{F814W}<21$), we counted 95 RGB stars in the HST dataset and
87 stars in the LBT dataset, thus suggesting a completeness lower limit of 90\%. Since the completeness is expected to further increase for larger clustercentric distances and for brighter stars, we conclude that the LBT catalogue provides an excellent completeness in the region not sampled by the HST observations and within the magnitude range considered in the analysis.}
Since the LBT filters are less efficient in separating MPs than the adopted HST ones (in particular at low metallicities, as in the case of NGC~2419) and the LBT photometric quality is lower than that provided by HST, in the following analysis we will consider only the FP and SP sub-groups and we will not attempt further splitting of the SPs. We used the stars in common between the HST and LBT catalogues to translate the MP selection criteria adopted in the former, into the photometric bands of the latter.
In doing so, we considered only HST stars having a probability of belonging to the FP or SP larger than $75\%$ (as predicted by the Gaussian mixture model), to minimize contamination between the two populations. Moreover, we restricted the analysis to high-quality photometry stars, by considering only stars in the wide-field catalogue having sharpness $\left| sh\right|\leq0.2$ and excluding all the sources located in the innermost $30\arcsec$ from the cluster center, where the severe crowding condition significantly decreases the photometric quality of ground-based images. We found that in the wide-field catalogue FP and SP stars appear to be nicely separated in the filter combination $(I,C_{uVI})$ where $C_{uVI}=(u-V)-(V-I)$ (left-hand panel of Figure~\ref{fig:cmdBeccari}). We then verticalized the color distribution of RGB stars in this filter combination and analysed the position of FP and SP stars in the verticalized distribution ($\Delta_{uVI}$). The verticalized distribution of the selected stars is shown in the right-hand panel of Figure~\ref{fig:cmdBeccari}. {As can be seen also in the histograms of the verticalized distributions, the 
 sub-populations thus selected appear to show an almost bi-modal distribution in $\Delta_{uVI}$ with a maximum separation around $\Delta_{uVI}=0.45$, with stars having $\Delta_{uVI}<0.45$ belonging to the FP stars and the remaining to the SP}. We then applied this separation to the LBT stars observed in the area outside the HST FoV.
In this way we obtain a complete list of FP and SP stars across the entire cluster extension. Specifically, stars within $1.3r_{hl}$ (where $r_{hl}\sim53.4\arcsec$ is the cluster half-light radius) are extracted from the HST data-set, while stars beyond this limit are extracted from the LBT data-set. In such a way, 1104 stars are associated to the FP, while 1352 are associated to the SP, yielding a global population ratio of $N_{FP}/N_{TOT}=0.45\pm0.02$, in very good agreement with the results by \citet{Zennaro2019} for the HST sample only.

\subsection{Multiple Populations fractions} 
\begin{figure}
\centering
\includegraphics[scale=0.4]{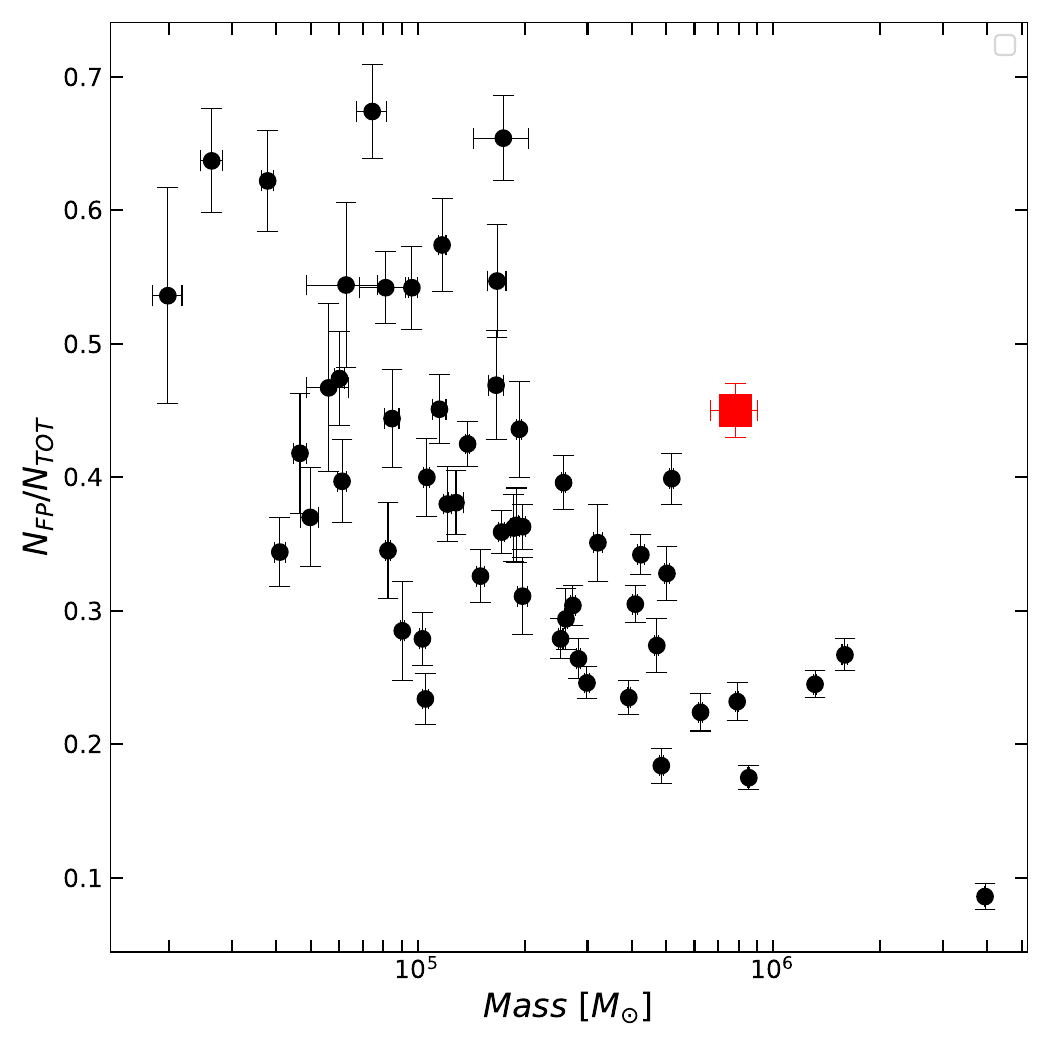}
\caption{Distribution of the observed fraction of FP stars (as obtained in \citealt{Milone2017} -- black circles) as a function of cluster mass (from \citealt{Baumgardt2018}). The red square shows the location of NGC~2419 based on the results found in this paper. }
\label{fig:massfrac}
\end{figure}

\begin{figure}
\centering
\includegraphics[scale=0.4]{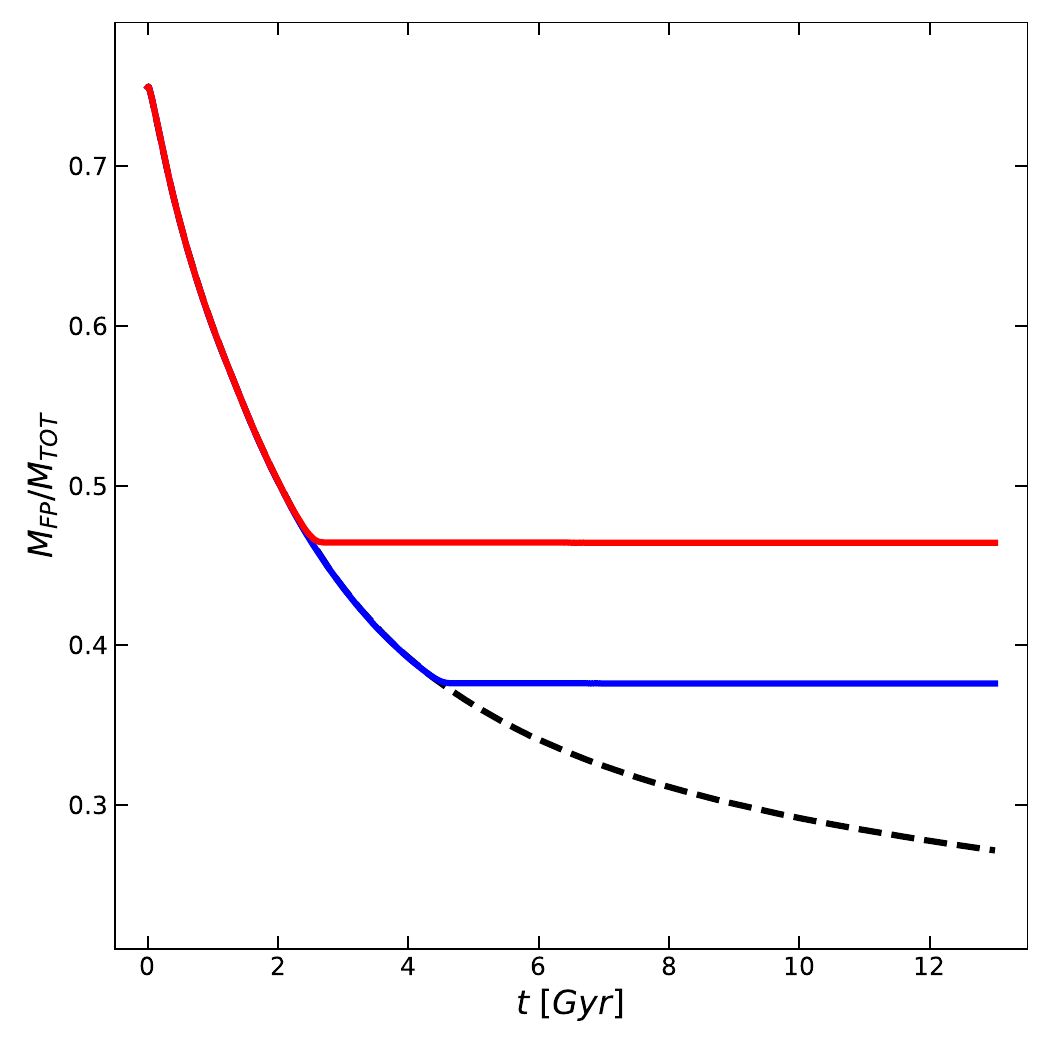}
\caption{Time evolution of the ratio of the FP mass to the total cluster mass as obtained by Monte Carlo simulations described in Section~3.
The black curve corresponds to the simulation in which a cluster was formed and spent its entire lifetime within a Sagittarius-like galaxy. 
The red and blue curves describe the $M_{FP}/M_{TOT}$ evolution if the cluster is accreated by a Milky Way like galaxy $2-3$ Gyr and $4-5$ Gyr after its formation, respectively.}
\label{fig:sims}
\end{figure}

Figure~\ref{fig:massfrac} shows the distribution of the FP stars number fraction ($N_{FP}/N_{TOT}$) (from \citealt{Milone2017})  
as a function of the present-day stellar mass obtained from \citealt{Baumgardt2018} for a sample of 53 Galactic GCs. 
As already discussed in \citet{Milone2017}, $N_{FP}/N_{TOT}$ decreases for increasing cluster mass. However, NGC 2419 appears to 
not follow the general trend, showing a significantly larger value than any cluster of comparable mass: indeed, it has the largest $N_{FP}/N_{TOT}$ ratio among the massive ($M> 5\times10^5 M_{\odot}$) 
Galactic GCs (see also \citealt{Zennaro2019}). A number of dynamical models of the evolution of $N_{FP}/N_{TOT}$ 
indicate that the strongest decrease of this ratio occurs during the GC early evolutionary phases 
(see e.g. \citealt{dercole08,Vesperini2021,Sollima2021}) and is driven by a significant early loss of FP stars, while the subsequent mass loss driven by two-body relaxation has a much weaker effect on the evolution of this ratio (see \citealt{Vesperini2021,Sollima2021}). 
A possible explanation for the observed value of $N_{FP}/N_{TOT}$ in NGC~2419 may be connected to the possibility that this cluster formed in the Sagittarius dwarf galaxy where, during its early evolutionary phases, it experienced a stronger tidal field than the one it is experiencing now in the very outer regions of the Milky Way.
Fig.~\ref{fig:sims} illustrates the time evolution of the ratio between the FP mass and the total cluster mass obtained from a Monte Carlo simulation run with the MOCCA code \citep{hypki13,giersz13}. 
This simulation starts with $6.5 \times 10^6$ stars with masses following a \citet{Kroupa2001} IMF between 0.1 and 100 $M_{\odot}$, and a ratio of the FP to total mass, $M_{FP}/M_{TOT}=0.75$. 
The SP is initially modeled as a King model with $W_0=7$ and a half-mass radius equal to 1/5 of 
the half-mass radius of the FP which initially follows the density profile of a King model with $W_0=4$. 
In the Monte Carlo simulation run for this paper, the system is initially assumed to evolve in a stronger tidal field than at its present-day galactocentric distance ($R_{gc}=95$ Kpc) in the Milky Way. Specifically the initial tidal field is equivalent, for example, to that at a distance of 2.5 kpc (4 kpc) from the center of a host dwarf galaxy modeled with a logarithmic potential with a circular velocity $V_c=50$ km/s ($80$ km/s) 
and with truncation radius of the FP coinciding with the tidal radius. We then assume that between 2 Gyr and 3 Gyr after the system formation, it is accreted in the Milky Way and the cluster continues its subsequent evolution at 95 kpc from the center of it. A similar simulation setup but with a transition to the Milky Way occurring between 4 Gyr and 5 Gyr has also been run.  As shown in Fig.~\ref{fig:sims}, until the transition to the weaker tidal occurs, $M_{FP}/M_{TOT}$ undergoes a significant decrease and follows the typical behavior found in previous studies \citep{Vesperini2021,Sollima2021}. After the transition to the much weaker tidal field at a Galactocentric distance of 95 kpc, the tidal radius is significantly larger than the cluster's size and the star loss rate becomes much weaker, essentially halting the evolution of $M_{FP}/M_{TOT}$. As shown by the dashed line, without such a transition $M_{FP}/M_{TOT}$ would continue to slightly decrease over the next 9-10 Gyr, eventually reaching ratios of about 0.2-0.3, compatible with those measured for the high-mass clusters shown in Figure~\ref{fig:massfrac}.

We strongly emphasize that the goal of this simulation is not to build a detailed model of NGC~2419 and its structural properties but rather to provide a simple illustration of how a possible transition to a much weaker tidal field than the one in which the cluster formed might reduce  the decrease of $M_{FP}/M_{TOT}$ and explain the high ratio of $M_{FP}/M_{TOT}$ found in NGC~2419 compared to other clusters with similar masses.

\section{MPs structural properties and radial distributions}
\subsection{Density profiles}
\begin{figure}
\centering
\includegraphics[scale=0.38]{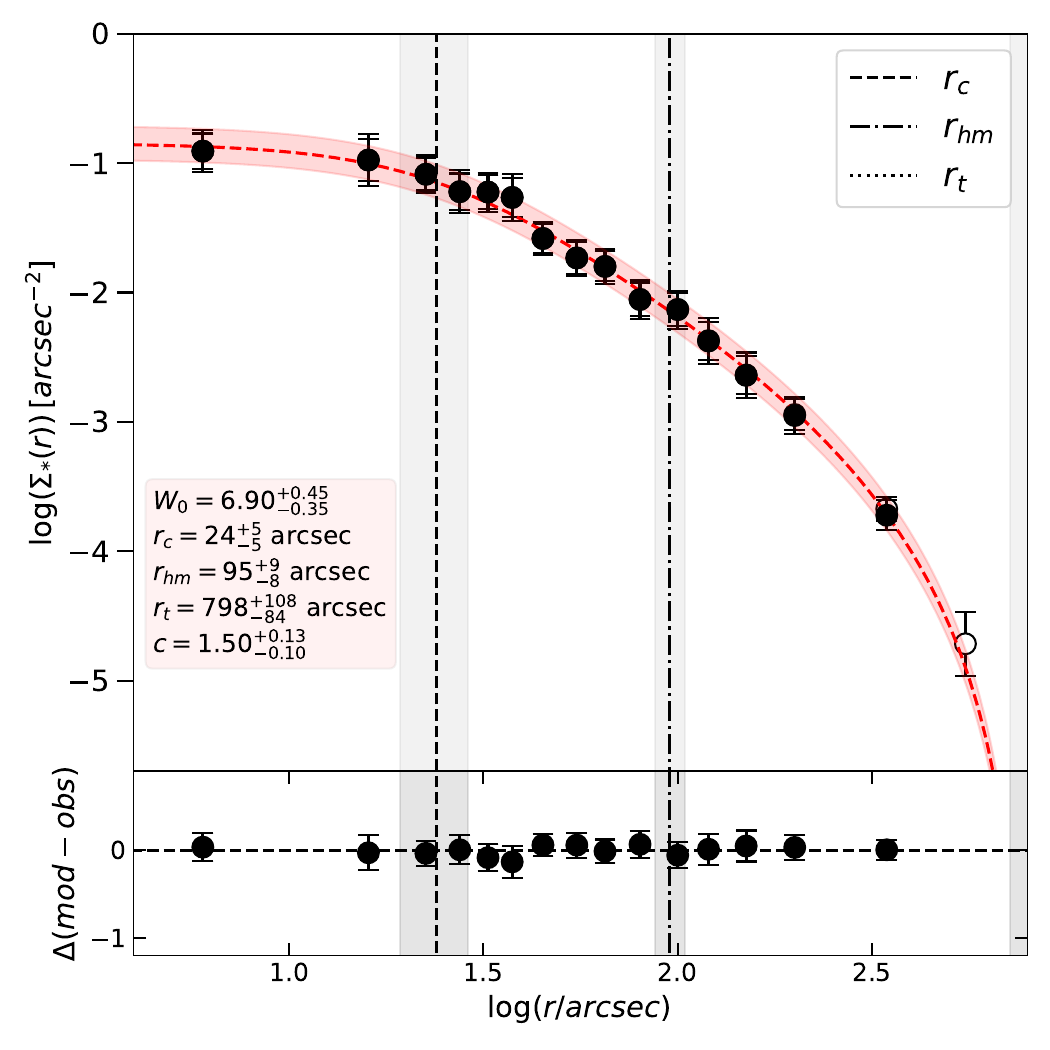}
\includegraphics[scale=0.38]{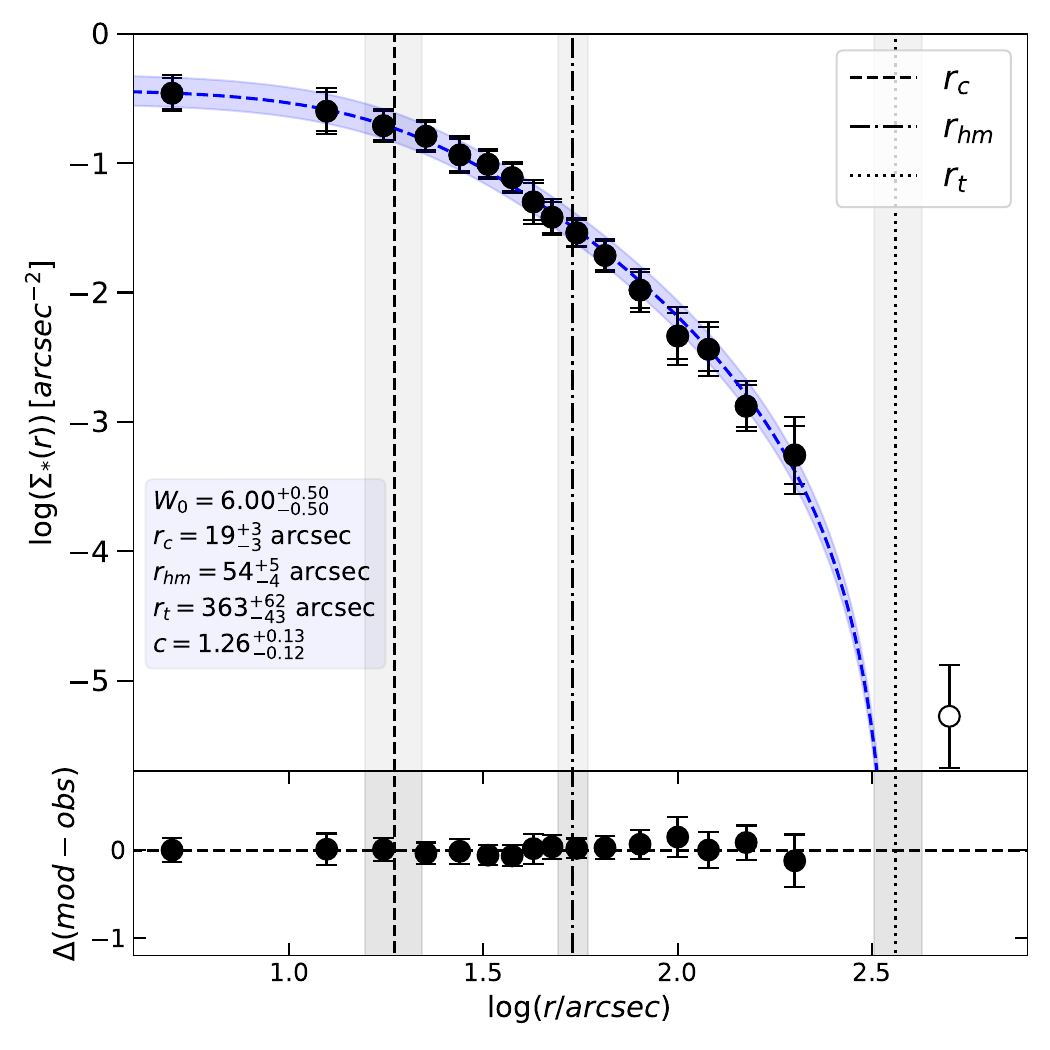}
\caption{ Stellar density profiles of MPs in NGC~2419. {\it Top panel:} Density profile of FP stars. Empty and solid circles mark, the observed and decontaminated density profiles of FP stars, respectively. The red dashed curve is the best-fit profile with a King model and the red stripe marks the envelope of the $\pm1\sigma$ solutions. The dashed, dot-dashed and dotted vertical lines mark the best-fit cluster’s core, half-mass and tidal radii, respectively, and their corresponding $1\sigma$ uncertainties are represented with the gray stripes. The fit residuals are plotted in the bottom panel. The best-fit structural parameters are reported in the inset box. {\it Bottom panel:} same as in top panel but for SP stars.}
\label{fig:densprof}
\end{figure}

 We obtained the density profiles of FP and SP stars following the prescriptions by \citet[][see also \citealt{cadelano17}]{lanzoni19}. Briefly, for each population, we divided the FoV in concentric annuli centered on the cluster gravitational center quoted by \citet{Dalessandro2008}. Each annulus was divided in sub-sectors in which we derived the mean stellar density and its standard deviation, adopted as uncertainty on the density value. The resulting density profiles for both the FP and SP are plotted in Figure~\ref{fig:densprof} (empty black circles). In both profiles the outermost value, which is located beyond the cluster's tidal radius ($r_t=460\arcsec$; \citealt{Dalessandro2008}), was assumed to be representative of the field background density. It was then subtracted to the other bins to obtain the decontaminated density profiles (filled black circles in Figure~\ref{fig:densprof}). 
 The decontaminated density profiles were then fitted using spherical, isotropic and single-mass \citet{King1966} models to derive the structural parameters of the two populations. Details on the fitting procedure are provided by \citet{raso20}. The best-fit curves and structural parameters are reported in Figure~\ref{fig:densprof}. The results of the fit clearly show that the two populations are characterized by significantly different structural properties. 
 In fact, despite being fit by King models with similar values of the central dimensionless potential $W_0$, the SP is characterized by a smaller core, half-mass and truncation radii than the FP. The ratios between the core, half-mass and truncation radii of the FP and SP are $r_c^{\rm FP}/r_c^{\rm SP}=1.3\pm0.3$, $r_{hm}^{\rm FP}/r_{hm}^{\rm SP}=1.8\pm0.2$ and $r_{t}^{\rm FP}/r_{t}^{\rm SP}=2.2\pm0.5$, respectively, thus confirming that the SP is more centrally concentrated than the FP. We do not find significant variations in the final results if slightly different values of $\Delta_{U,V,I}$ are used to separate between the two populations in the LBT data-set.
 We note that \citet{larsen19} found a significantly smaller difference between the half-mass radii of the two sub-populations, with the SP half-mass radius being only $\sim10\%$ smaller than the FP one. This is likely due to the limited FoV adopted in their analysis, which only sample a radial range of 0.7-0.8 $r_{hm}$. 


\begin{figure}
\centering
\includegraphics[scale=0.35]{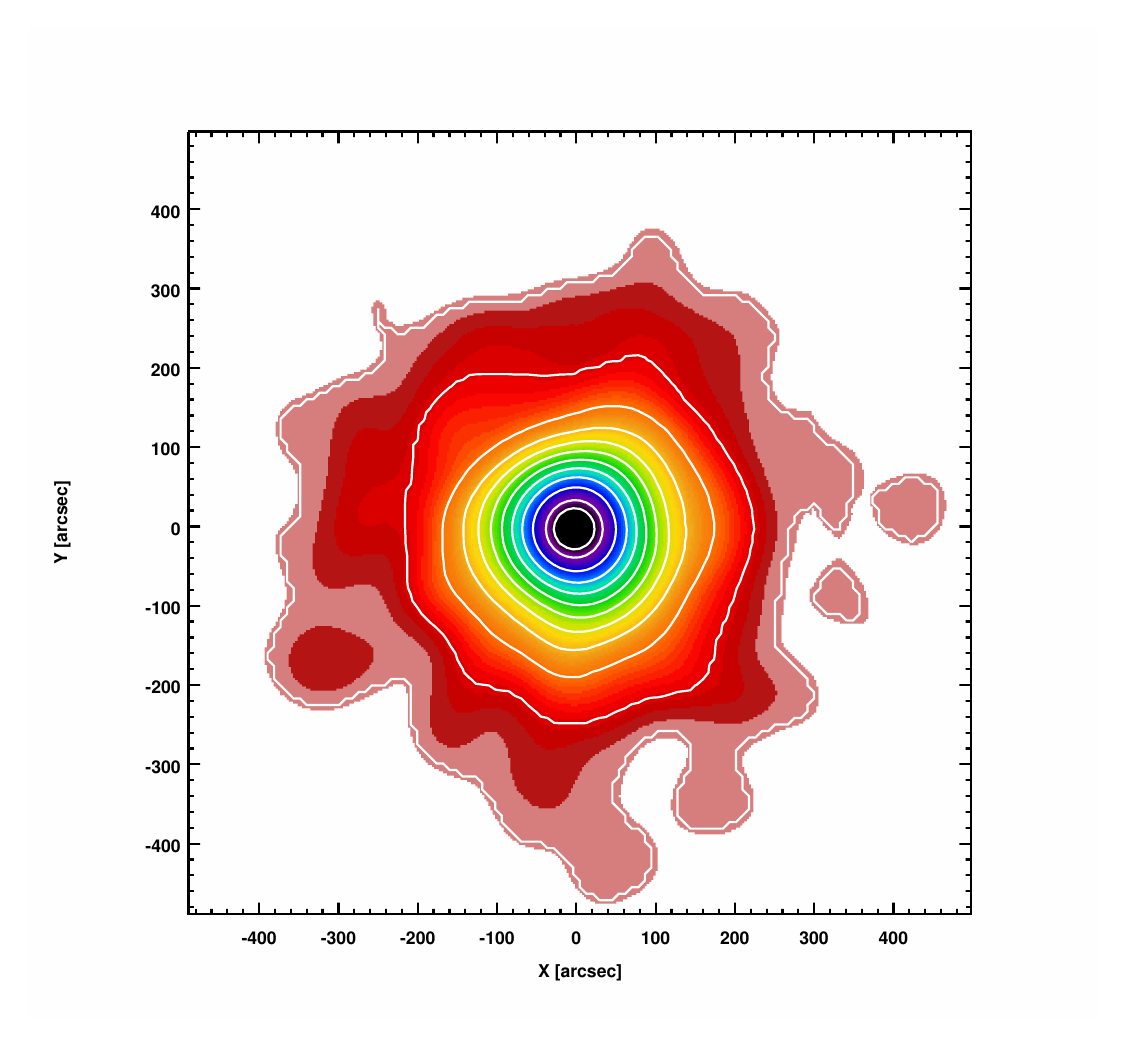}
\includegraphics[scale=0.35]{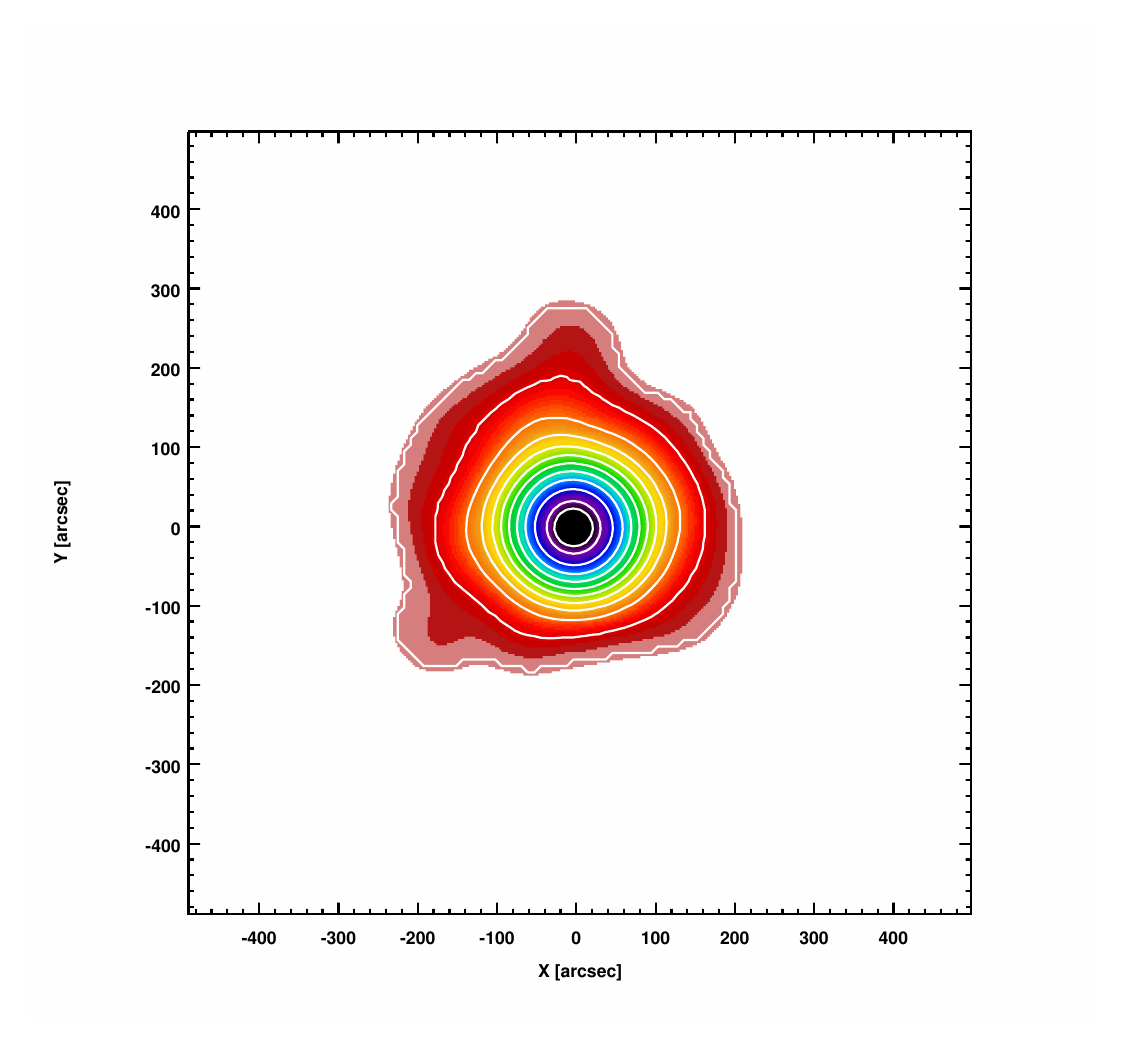}
\caption{2D density maps and isodensity contours of the FP and SP in the top and bottom panels respectively.}

\label{fig:densmap}
\end{figure}

 In addition to the projected density profiles, we explored also the 2D density distributions for both the FP and SP. The 2D density maps shown in Figure~\ref{fig:densmap} were obtained by transforming the sky distribution of the selected RGB stars into a smoothed surface density function through the use of a Gaussian kernel with width of $20\arcsec$.  Also the 2D maps clearly show that the SP is more centrally concentrated than the FP. Overall, both distributions appear to be grossly spherical, in agreement with the low ellipticity ($\epsilon=0.03$) derived for NGC~2419 \citep{Harris2010}. However, we note that the SP density distribution is slightly more elongated than the FP one and it reaches a maximum $\epsilon\sim0.08$, with the ellipse major axis directed along the East-West direction. This might be the effect of a rotating SP.

\begin{figure}
\centering
\includegraphics[scale=0.23]{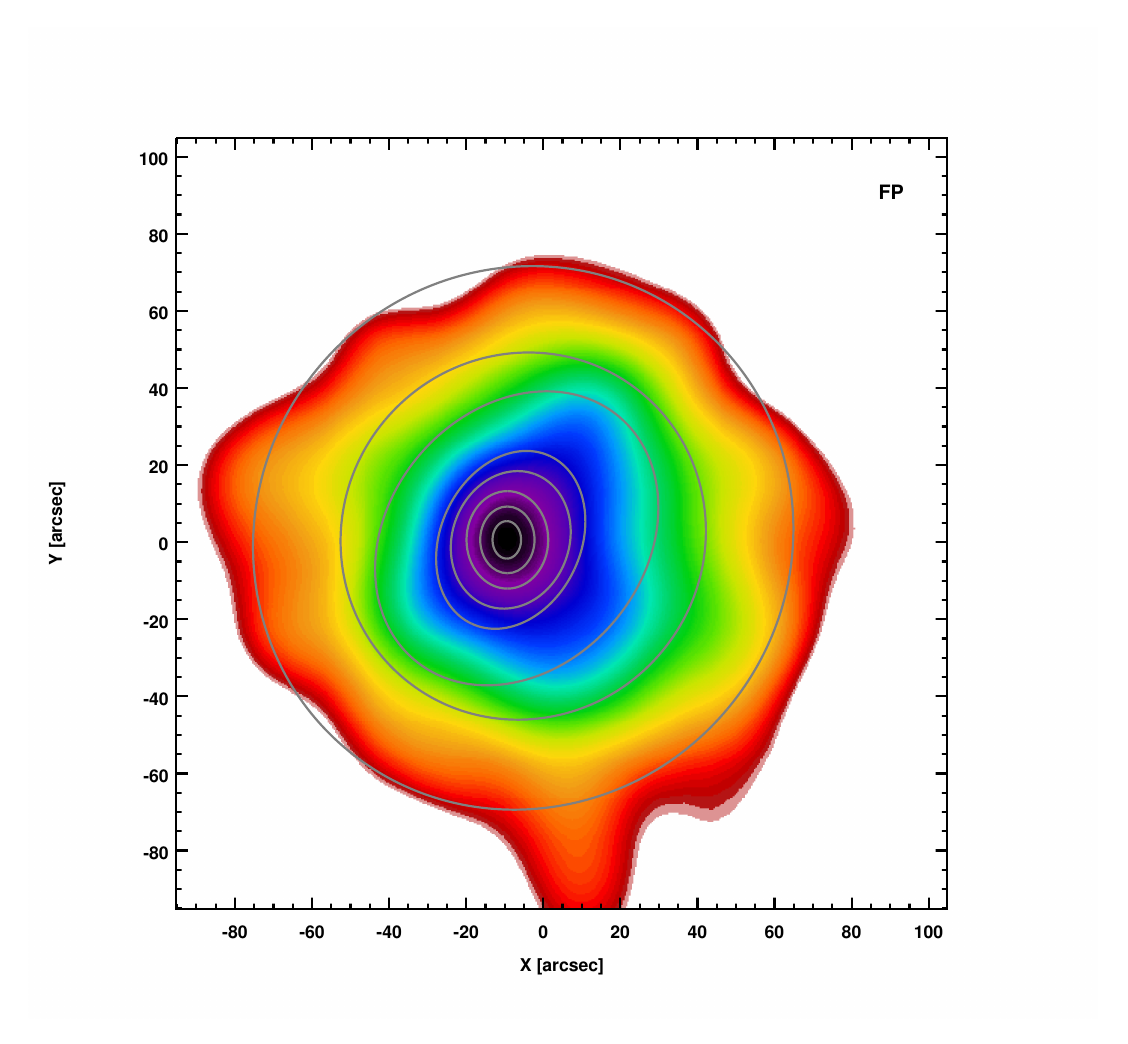}
\includegraphics[scale=0.23]{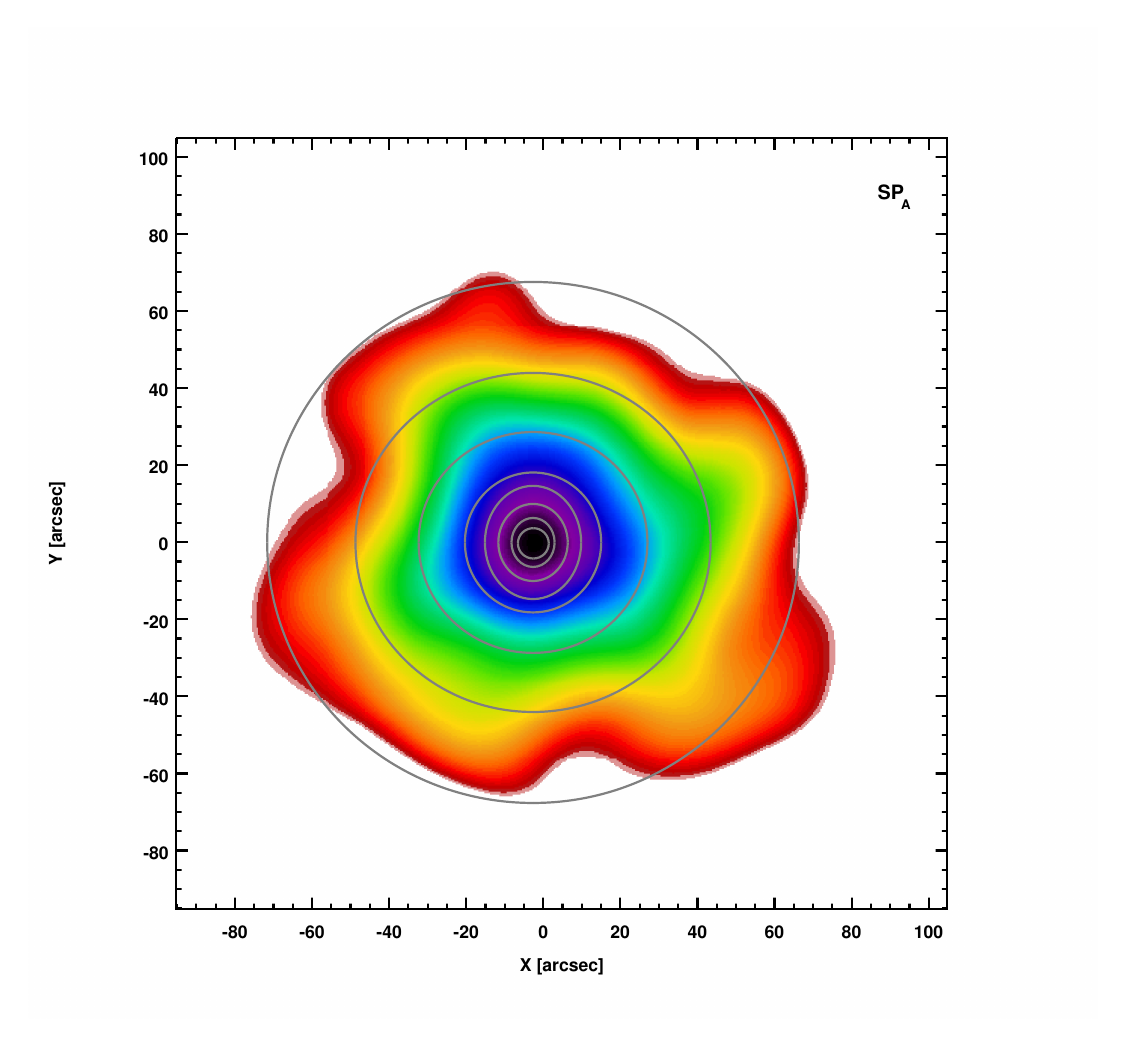}
\includegraphics[scale=0.23]{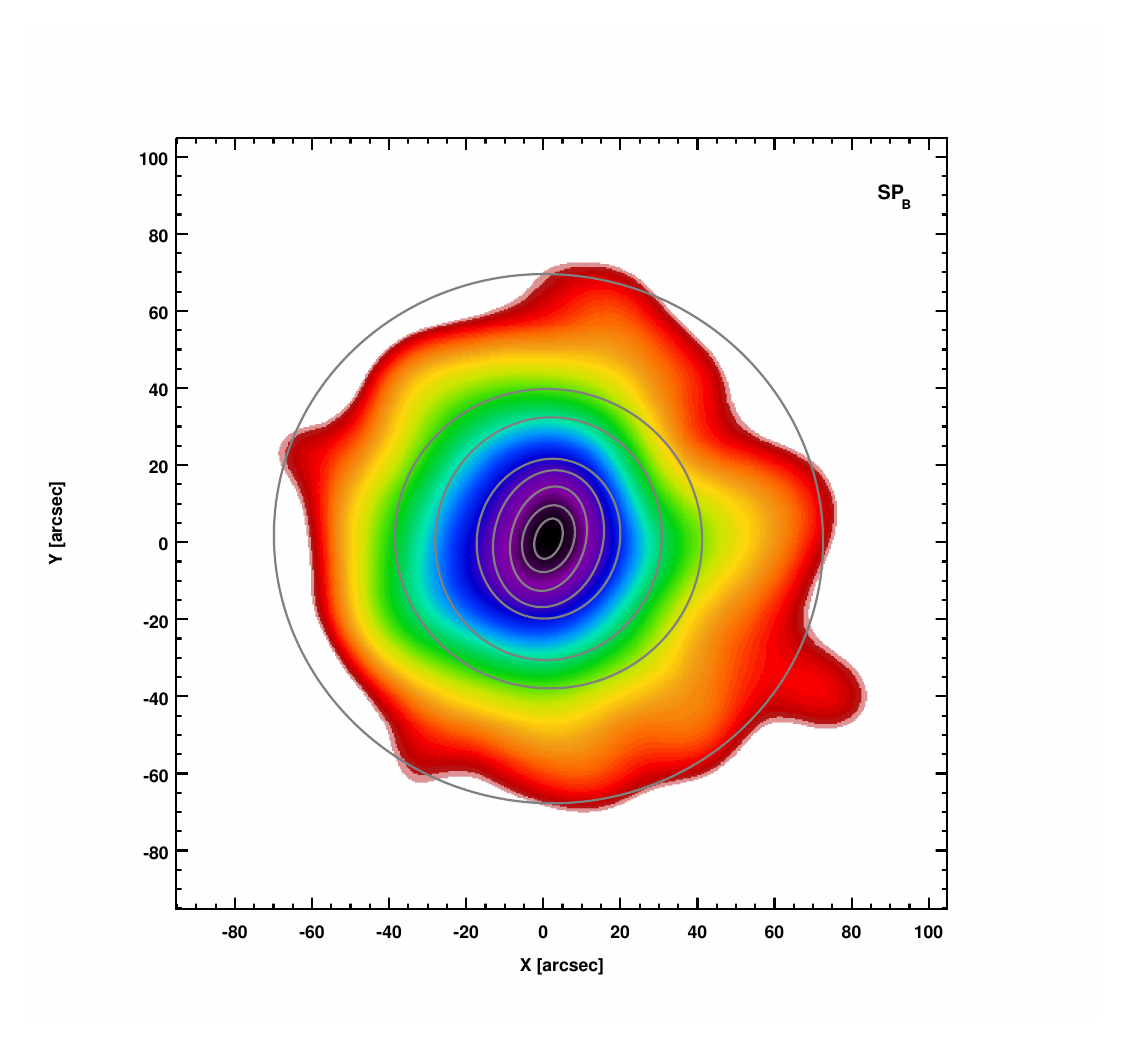}
\includegraphics[scale=0.23]{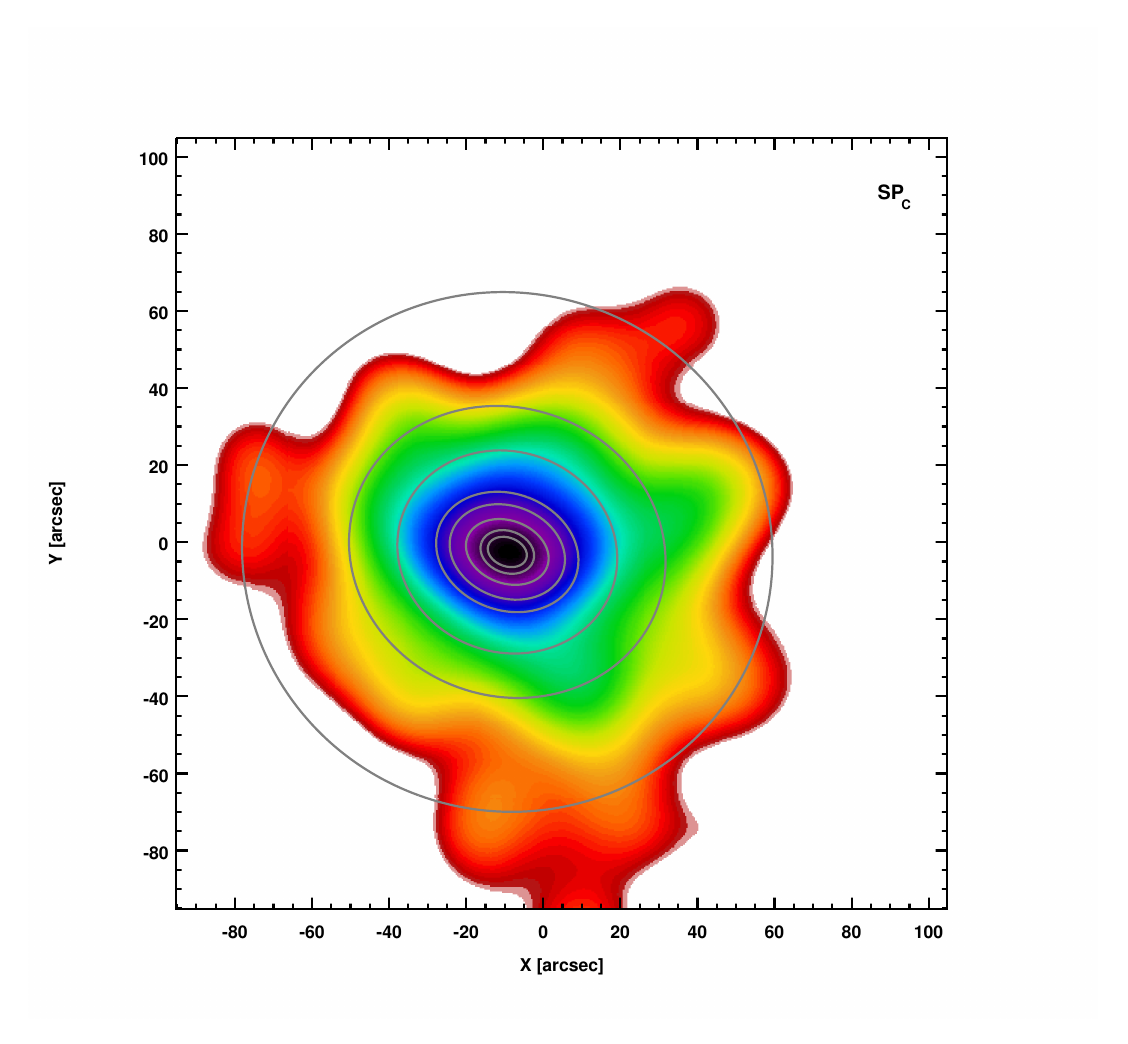}

\caption{2D density maps and isodensity contours for the HST stars and belonging to the FP (left-hand top panel), SP$_A$ (right-hand top panel), SP$_B$ (left-hand bottom panel) and SP$_C$ (right-hand bottom panel).}
\label{fig:densmaphst}
\end{figure}

\begin{figure}
\centering
\includegraphics[scale=0.3]{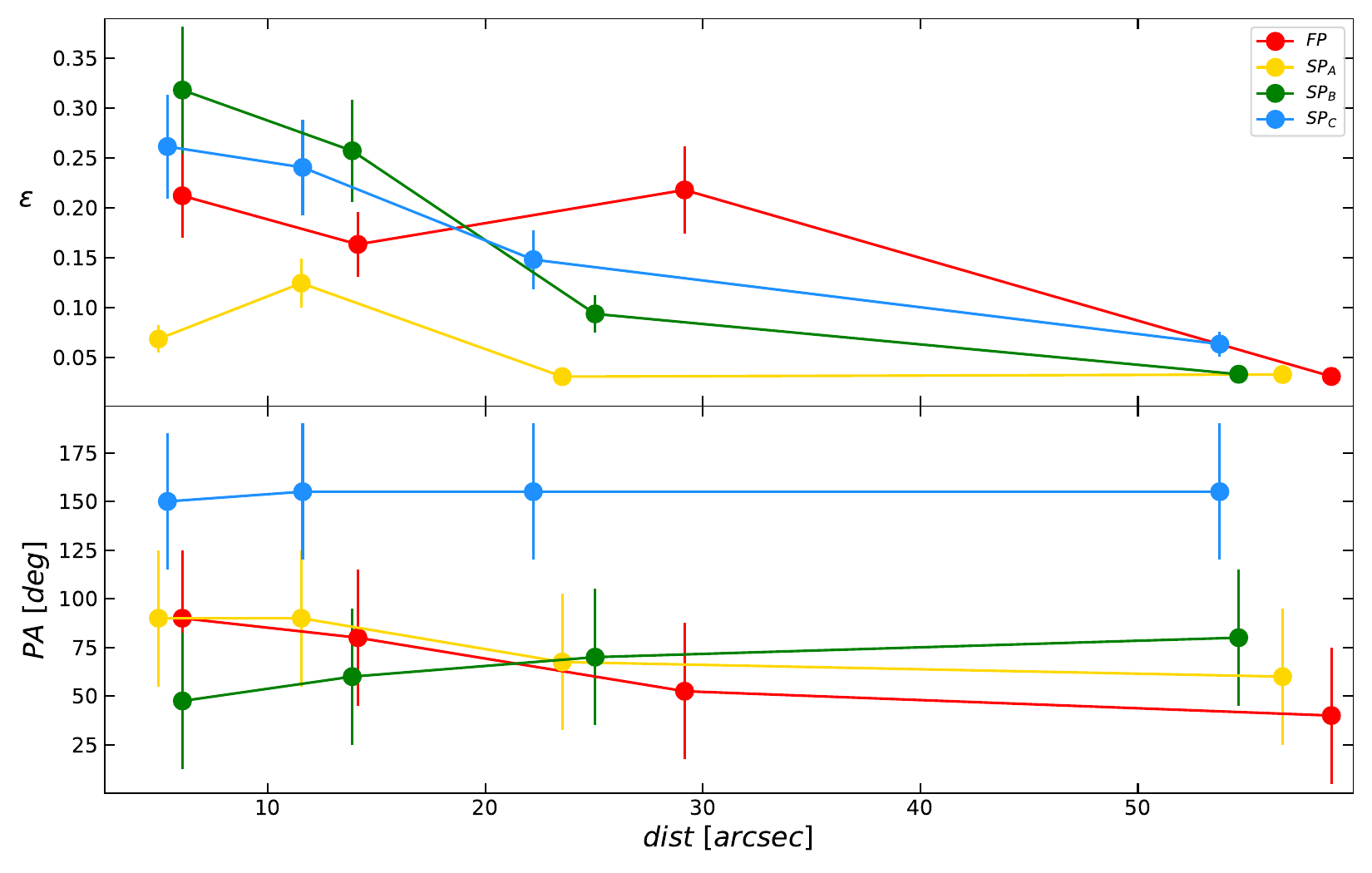}
\caption{Ellipticity and position angle (top and bottom panel, respectively) of the isodensity contours shown in Figure~\ref{fig:densmaphst} for the different sub populations, as a function of the radial distance from the cluster center. The angles are measured counterclockwise with the origin set along the east direction.}
\label{fig:ellpa}
\end{figure}

We further investigated the 2D density distribution of the FP, SP$_A$, SP$_B$, and SP$_C$ sub-populations identified within the HST FoV, {using instead a Gaussian kernel with a width of $5\arcsec$}. The maps plotted in Figure~\ref{fig:densmaphst} show that, with the only exception of SP$_A$, all the populations are characterized by elongated elliptical structures in the innermost cluster regions. As shown in the top-panel of Figure~\ref{fig:ellpa}, they have an ellipticity of about  $\epsilon\sim0.2-0.3$ within the cluster core radius ($r_c=20\arcsec$). This value then decreases to zero beyond the core, with the exception of the FP. The only exceptions are the SP$_A$, which shows a lower ellipticity around 0.1, and the FP, which retain a significant ellipticity within twice $r_c$. The ellipticity of SP$_A$, instead, is always smaller than the others, never exceeding $\epsilon\sim0.15$ and rapidly vanishing to zero. The bottom panel of Figure~\ref{fig:ellpa} shows the position angle (PA) of the isodensity contours (the angle is measured counterclockwise with the origin set in the east direction). Interestingly, while the FP and SP$_{A,B}$ show a constant and similar position angle of about $75\deg$ throughout the whole HST FoV, the SP$_C$ shows a remarkably different position angle of about $150\deg$, which is almost perpendicular with respect to the other three populations. This intriguing feature suggests that the SP$_c$ population may be characterized by a very different kinematic. Since this cluster is dynamically young, such a difference may contain some memory of the properties emerging from the formation and early evolutionary phases.
However, given the low statistic of the sample, additional observations are needed to suitably investigate the kinematical differences of the various SPs.

\subsection{Radial distribution of Multiple Populations}
\begin{figure}
 \centering
 \includegraphics[scale=0.35]{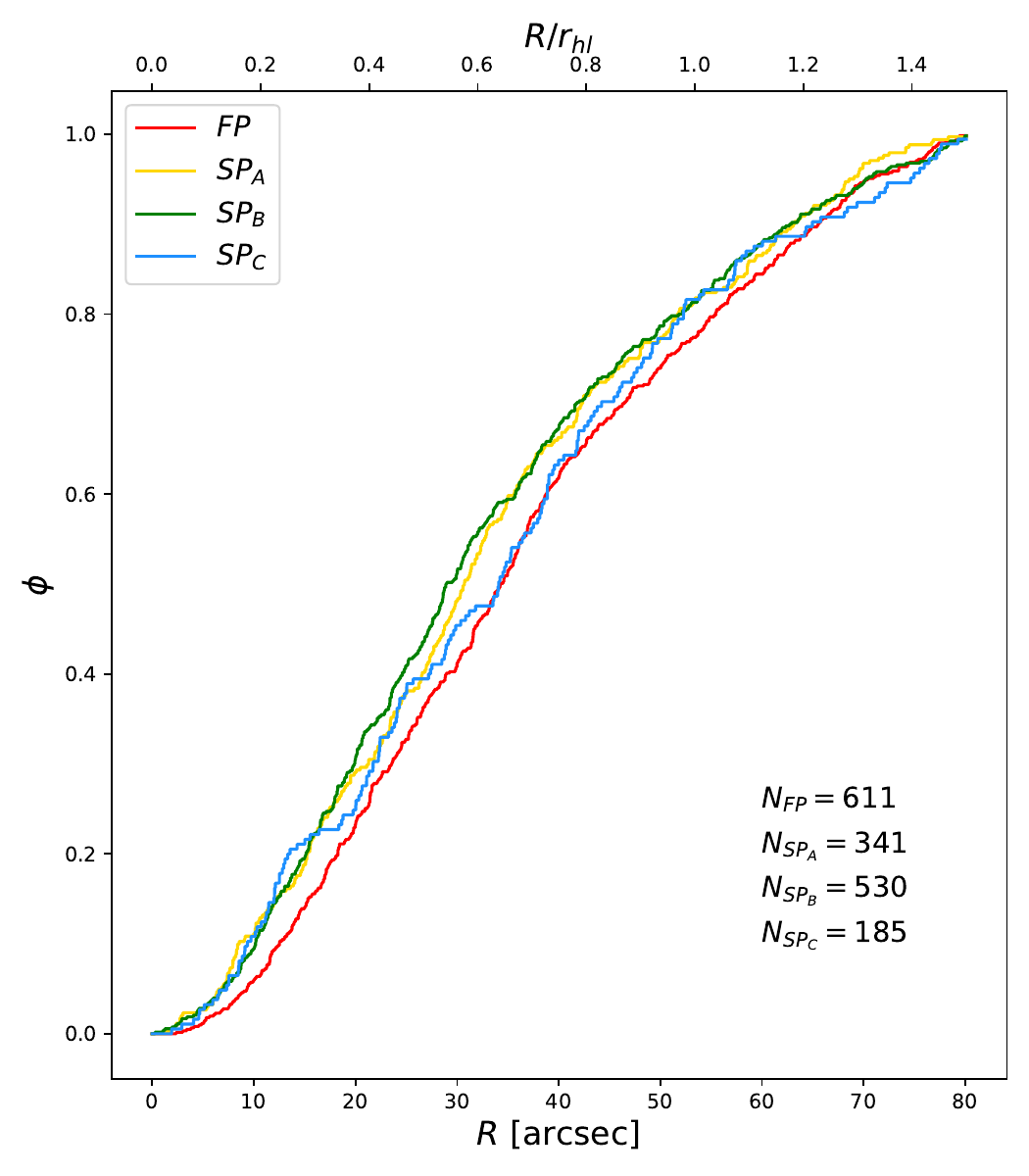}
 \caption{Cumulative radial distributions of the four sub-populations of NGC 2419 detected  within the HST FoV (the color-code is the same used in Figure~\ref{fig:chm}).}
 \label{fig:crdHST}
\end{figure}

We can now compare the radial distributions of FP and SP stars. 
In Figure~\ref{fig:crdHST} we show the cumulative radial distributions of the 4 sub-populations identified in the HST data-set. They are presented for illustrative purpose only, as they sample a limited cluster extension ($\sim1.5 r_h$) and therefore may be not representative of the global behavior.
As expected, all SP sub-populations tend to be more centrally concentrated than the FP one. This result is particularly clear in the case of SP$_A$ and SP$_B$ stars, while for SP$_C$ is far less significant. Although this could be another signature of a different formation and evolutionary path of SP$_C$ stars, we stress that both the low number statistic for this sub-population and the limited radial coverage of the HST data-set do not allow us to further elaborate on the possible different properties of the SP$_C$ population.

The general behavior of the SP radial distributions within the HST FoV is qualitatively in agreement with that presented by \citet{larsen19}, although there is not a one-to-one correspondence between the sub-populations defined here and those defined by \citet{larsen19} due to the different adopted selection methods.

\begin{figure}
 \centering
  \includegraphics[scale=0.3]{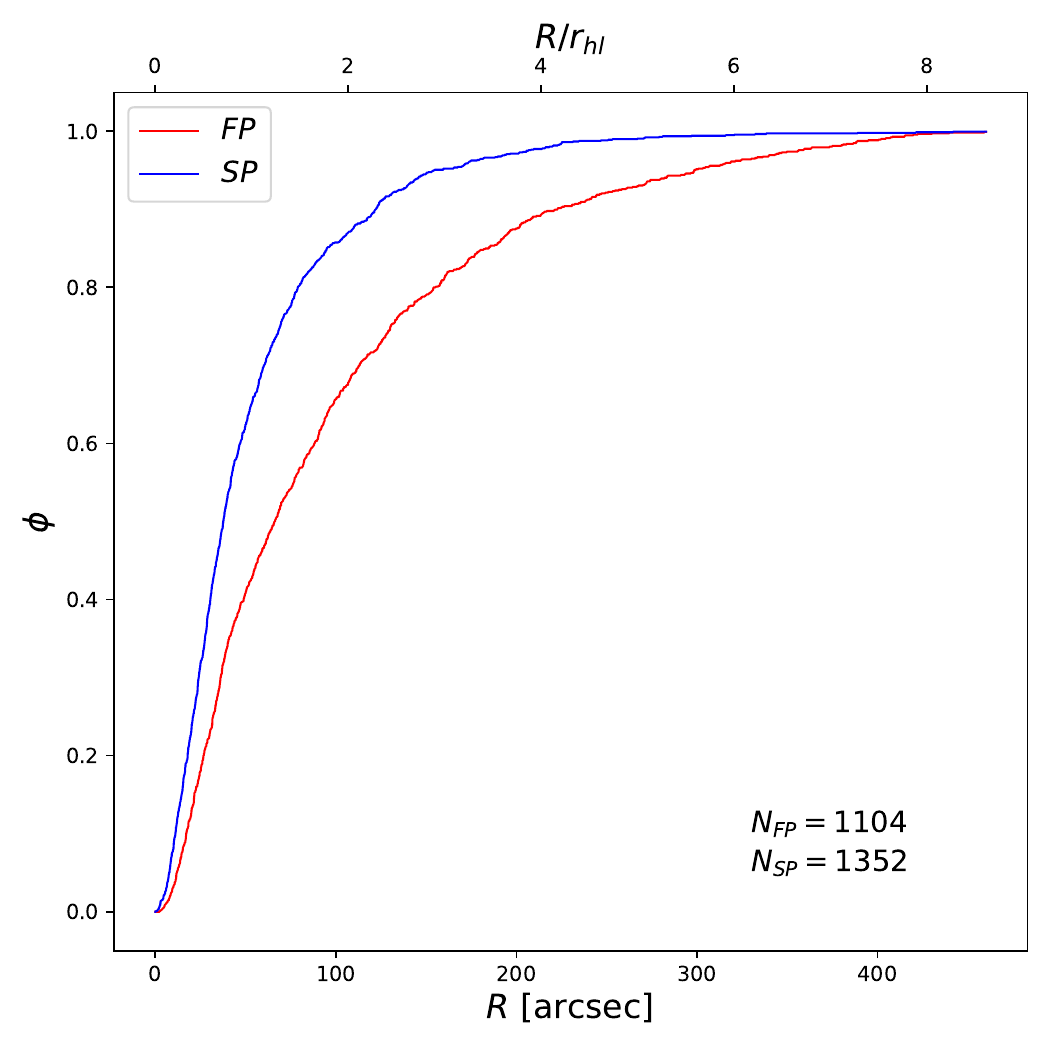}
   \includegraphics[scale=0.3]{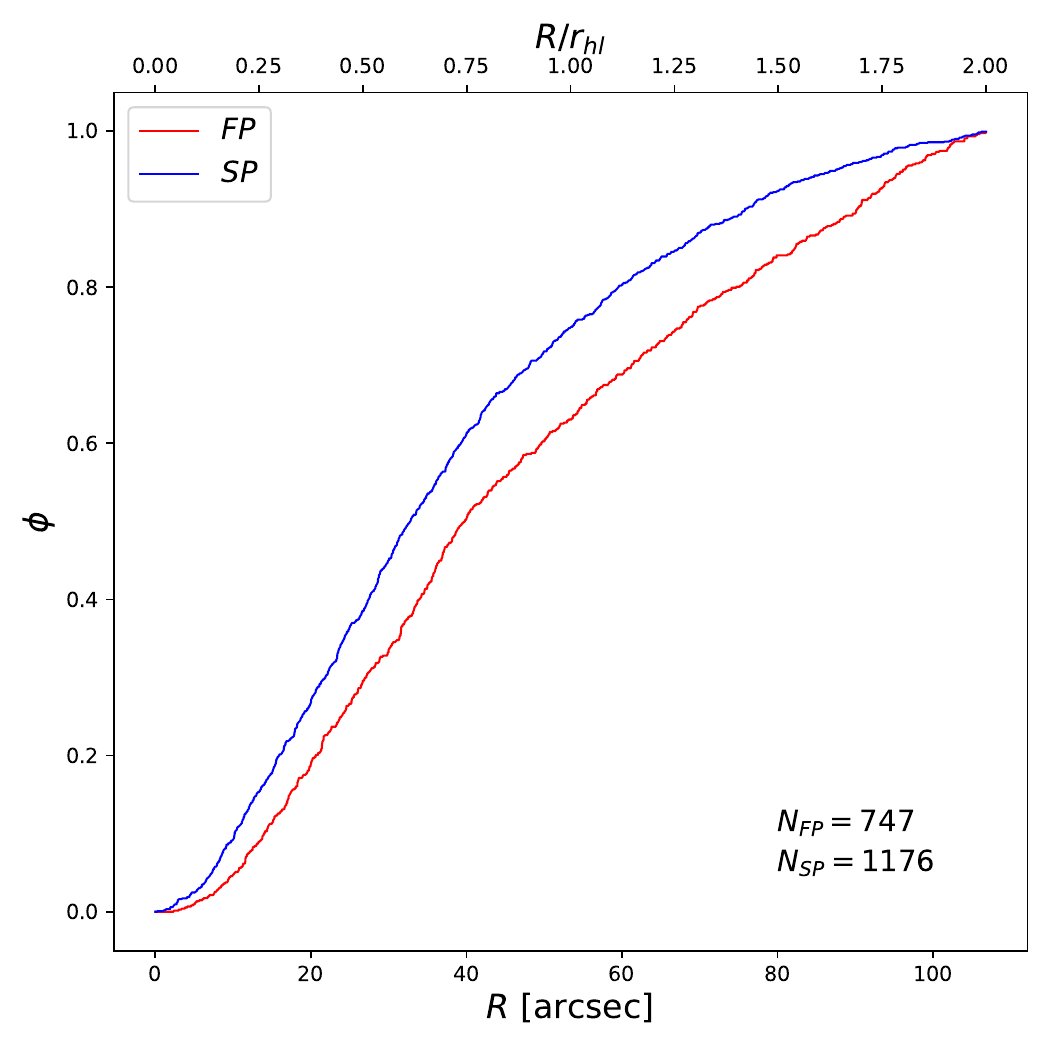}
  \caption{Cumulative radial distribution of FP stars (red curve) and SP stars (blue curve) obtained combining together the HST and LBT data-set. The top panel shows the distribution of stars across the whole cluster extension while the bottom panel shows the distribution within twice the cluster half-light radius.}

 \label{fig:crdALL}
\end{figure}

As discussed above, the study of the MP properties along the entire radial extension of the cluster is limited to the two sub-groups of FP and SP stars as obtained from the combination of the HST and LBT data-sets. The left-hand panel of Figure~\ref{fig:crdALL} shows the cumulative radial distributions of the two populations out to the cluster's tidal radius. In agreement with results obtained from the density profile analysis, we find that the SP is significantly more centrally segregated than the FP one across the whole cluster extension. Indeed, the Kolmogorov-Smirov test basically returns a null probability ($P_{KS}\sim10^{-31}$) that the two samples are drawn from the same distribution!

\begin{figure}
 \centering
 \includegraphics[scale=0.28]{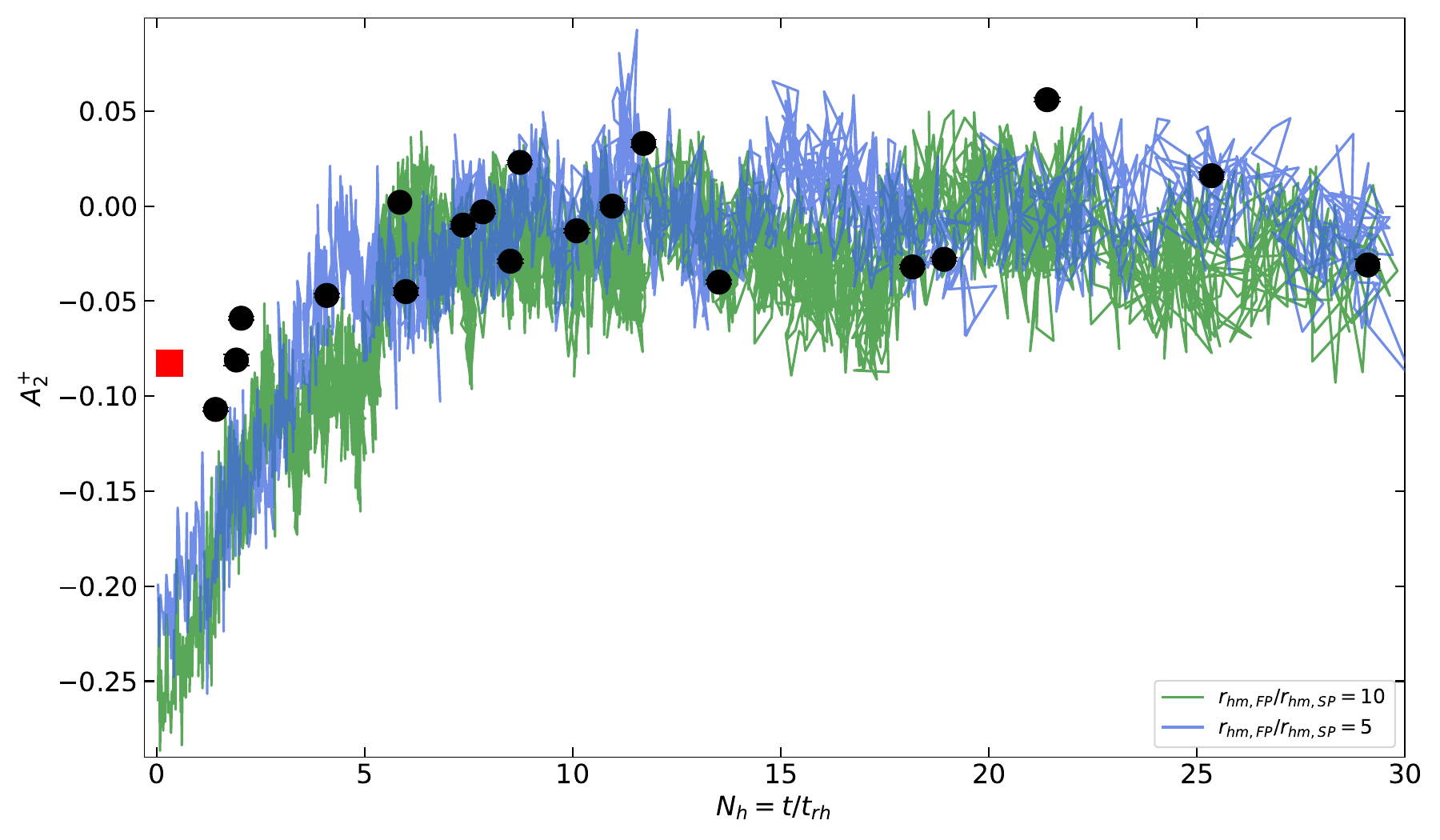}
 \caption{$A^+_2$ parameter as a function of the dynamical age $N_h$. The black circles are the GCs analysed in \citet[][see also \citealt{Smolinski2020}]{Dalessandro2019}, the red square is NGC~2419. The blue and green lines show the results of the $N$-body simulations presented by \citet{Dalessandro2019} following the $A^+_2$ evolution of clusters born with a FP half-mass radius 5 and 10 times larger than that of SP stars, respectively.}
 \label{fig:apiu}
\end{figure}

The available data-set allows us to put NGC~2419 in the framework of MP dynamical evolution presented by \citet{Dalessandro2019} and discussed in the Introduction. 
We measured the $A^+$ parameter only for stars within twice the cluster half-light radius ($A^+_2$). 
We show the cumulative distributions of FP and SP stars within $2r_{hl}$ in the right-hand panel of Figure~\ref{fig:crdALL}. The area between the two curves provided us with the value of the $A^+_2$ parameter, while to quantify its uncertainty we performed a bootstrapping/jackknife re-sampling, finally finding  $A^+_2=-0.0826\pm0.0005$. 
Assuming a cluster age $t=13$ Gyr and $t_{rh}=42.7$ Gyr \citep{Dotter2008,Harris2010} we find that the ratio $N_h=t/t_{rh}$, which can be used as a measure of the cluster dynamical age, is of only $N_h\sim0.3$. The position of NGC~2419 in a diagram showing $A_2^+$ as a function of $N_h$ is plotted in Figure~\ref{fig:apiu}, compared with the results of N-body simulations (green and blue lines) and the location of the GCs analysed by \citet{Dalessandro2019}, with the addition of M~13 from \citealt{Smolinski2020}. 
The value of $A^+_2$ obtained for NGC~2419 is among the most negative in the sample and it quite nicely follows the overall trend in which less dynamically evolved clusters (t/t$_{rh} < 8 - 10$) have SP stars more centrally concentrated than the FP ones (i.e. negative values of $A^+$).
%
This trend is consistent with that expected from models of dynamical evolution and spatial mixing of multiple-population clusters. In order to illustrate the expected trend we show in the same plot the results of the two N-body simulations presented in \citet{Vesperini2018} and already discussed also in \citet{Dalessandro2019}. The results of these two simulations are added just to demonstrate the expected trend but a larger survey of models, providing a comprehensive coverage of the various initial dynamical properties of MPs (such as the initial ratio of FP to SP half-mass radii, the initial FP and SP density profiles, as determined, for example, by the King central dimensionless potential $W_0$, the FP and SP kinematics) would be necessary to establish a closer connection between the theoretical predictions and observations, and to shed light on the possible evolutionary path leading to the present-day structure of NGC~2419.
  
\section{Conclusions}

In this paper we have presented the detailed characterization of the structural properties of MPs in NGC~2419, one of the most massive ($M\sim10^6 M_{\odot}$) and dynamically young ($t_{age}$/$t_{rh}\sim0.3$) GC in the Galaxy.

In our analysis we have combined data from HST and LBT to study the density profile and the morphology of MPs over a radial range extending from the cluster's center to the tidal radius $r_t$.
Our results show that the SP is more spatially concentrated than the FP one. The density profiles of the two populations can be fit by King models with similar values of the central dimensionless potential ($W_0=6.9$ for the FP and $W_0=6$ for the SP) but different spatial scales: the core and half-mass radius of the FP are respectively, about 1.3 and 1.8 times larger than those of the SP, while the ratio between the King truncation radius of the FP to the SP is equal to about 2.2.

2D maps of the spatial distributions of the MPs reveal some deviations from spherical symmetry which might be associated to internal rotation. Interestingly the heavily enriched $SP_c$ shows a remarkably different structure, being elongated in the central regions almost perpendicularly to the other sub-populations. Ad-hoc spectroscopic observations are necessary to shed light on the possible peculiar kinematics of this population which may provide clues on its formation process. 

We have quantified the differences in the spatial distribution of the FP and SP stars by means of the $A^+$ parameter and found that NGC~2419 has one of the most negative $A^+_2$ values for the sample of GCs in which this quantity has been estimated \citep{Dalessandro2019} and it qualitatively follows the expected general trend $A^+_2-t/t_{rh}$.

The characterization of the structural properties of MPs in such a dynamically young cluster provides an important empirical constraint on the properties emerging after the formation and early evolution phases (see e.g. \citealt{Vesperini2021,Sollima2021}) and it represents a key ingredient to inform models aimed at studying the long-term evolution driven by two-body relaxation.

We also confirm that, with a value of $\sim45\%$, the fraction of FP stars in NGC~2419 is significantly larger than what observed for clusters of similar mass (see also \citealt{Zennaro2019}).
A possible solution to this anomaly comes from the fact that the orbital properties of NGC~2419 suggests it was originally formed in the Sagittarius dwarf galaxy \citet{massari19}.
We have shown that a transition from the tidal field of the original host galaxy to the very weak tidal field at the present-day Galactocentric distance would significantly slow down the rate of star loss (preferentially affecting FP stars) and halt the evolutionary decrease of the FP fraction.

A tailored set of numerical simulations are needed to build models specifically aimed at reconstructing the dynamical history of NGC~2419.
The detailed observational characterization of the FP and SP structural properties of this study  provides important constraints for these models and, more in general, for the dynamical study of MP clusters.
On the observational side, an extension of such a detailed dynamical characterization to a larger sample of clusters and its combination with data on the internal kinematics is necessary to build a complete empirical dynamical characterization of multiple populations.

\begin{acknowledgements}
We kindly thank Giacomo Beccari for sharing the LBT photometric catalogue used in this work.
M.C. and E.D. acknowledge financial support from the project Light-on-Dark granted by MIUR through PRIN2017-2017K7REXT. E.D. acknowledges support from the Indiana University Institute for Advanced Study through the Visiting Fellowship program. E.V. acknowledges support from NSF grant AST-2009193.  
\end{acknowledgements}


\bibliography{bib}

\end{document}